\documentclass[fleqn,10pt]{wlscirep}
\usepackage[utf8]{inputenc}
\usepackage[T1]{fontenc}
\usepackage{graphicx}
\graphicspath{{./figures/}}
\DeclareGraphicsExtensions{.pdf,.jpeg,.png}
\usepackage{epstopdf}
\usepackage{subfig}
\usepackage{floatrow}
\usepackage{color}
\usepackage{multirow}
\usepackage{longtable}
\usepackage{comment}
\newcommand{\beginsupplement}{
        \setcounter{table}{0}
        \renewcommand{\thetable}{S\arabic{table}}
        \setcounter{figure}{0}
        \renewcommand{\thefigure}{S\arabic{figure}}}
        
\title{Sentiment Correlation in Financial News Networks and Associated Market Movements}

\author[1*$\dagger$]{Xingchen Wan}
\author[2,3*$\dagger$]{Jie Yang}
\author[4]{Slavi Marinov}
\author[1]{Jan-Peter Calliess}
\author[1]{Stefan Zohren}
\author[1]{Xiaowen Dong}

\affil[1]{Oxford-Man Institute of Quantitative Finance, University of Oxford, UK}
\affil[2]{School of Public Health, Zhejiang University,  China}
\affil[3]{Harvard Medical School, Harvard University, USA}
\affil[4]{Man AHL, UK} 
\affil[*]{xwan@robots.ox.ac.uk, jieynlp@gmail.com}
\affil[$\dagger$]{these authors contributed equally to this work.}

\begin{abstract}
In an increasingly connected global market, news sentiment towards one company may not only indicate its own market performance, but can also be associated with a broader movement on the sentiment and performance of other companies from the same or even different sectors. In this paper, we apply NLP techniques to understand news sentiment of 87 companies among the most reported on Reuters for a period of seven years. We investigate the propagation of such sentiment in company networks and evaluate the associated market movements in terms of stock price and volatility. Our results suggest that, in certain sectors, strong media sentiment towards one company may indicate a significant change in media sentiment towards related companies measured as neighbours in a financial network constructed from news co-occurrence. Furthermore, there exists a weak but statistically significant association between strong media sentiment and abnormal market return as well as volatility. 
Such an association is more significant at the level of individual companies, but nevertheless remains visible at the level of sectors or groups of companies.
\end{abstract}
\begin{document}

\flushbottom
\maketitle
%
%
\thispagestyle{empty}


\section*{Introduction}


Complexity and inter-dependencies have been the defining features of most modern financial markets: a myriad of ever-changing interactions between market participants, financial assets and relations with broader macroeconomic factors have all contributed to intricate market dynamics. Yet, with recent examples of failures of financial systems triggering broader economic avalanches \cite{longstaff2010subprime}, it has been both an empirical and an academic goal to untangle and shed new insights into the mechanisms of financial markets. 

While classical economic models have often been found inadequate in explaining the lower-level market dynamics \cite{battiston2016complexity}, there has been increasing interest and success, to varying degrees, in applying data-driven and computational modelling approaches in this context. One prominent example is the recent attempt to model financial markets as computational networks, with the individual components (for example, the individual stocks or other financial assets) being the nodes and the correlation or some other form of relation between these components being edges \cite{mantegna1999hierarchical,almog2019structural,onnela2003asset, heimo2009maximal}. This model and its variants have been used, for example, to model the propagation of systemic risk in wider economic systems at a global scale \cite{starnini2019interconnected} and to assess default risk propagation in sectorial networks \cite{barja2019assessing}. 

Concurrently, with the advancement of natural language processing (NLP) techniques \cite{opensent, lisentiment}, the algorithmic analysis of large-scale unstructured data that were previously not amenable to numerical processing (such as text data in financial news articles or research reports) has become possible. In this aspect, works have been dedicated to constructing networks of news and sentiment for various tasks, including sentiment analysis and risk/volatility monitoring  \cite{forss2018news, mizuno2017novel}. Nonetheless, to the best of our knowledge, there still is a lack of studies combining the study of the correlation of sentiment and the corresponding market movements: graph-based market dynamics analysis and NLP-based analysis on financial markets have been largely taken as independent directions so far -- while there have been studies on the market impact of sentiment \cite{si2013exploiting, mittal2012stock}, the analysis has been mostly restricted to the investigation on individual companies. This is clearly inadequate and sub-optimal: the evolution and collective dynamics of news and the sentiment network can be equally important, and non-trivial relations may exist between them that go beyond the individual company level.

In this work, we aim to bridge this gap by combining network analysis and NLP techniques to study sentiment correlation and its associated movements of the financial market. We focus on a collection of companies (which also contain many with the world's largest market capitalisation) that appear most frequently in news on Reuters. Specifically, we acquire news articles from Reuters on those companies of interest from 2007 to 2013: the period was deemed to span the late stages of pre-crisis boom, the 2008 Recession, the European debt crisis, and many important turns of events in the global markets. We apply NLP techniques to distill sentiment information and construct a network of 87 target companies based on the news information. We then perform community detection on this network and find that the groups of highly related companies obtained are often in close agreement with the conventional sectorial classification. 

The news occurrence network allows us to study correlation dynamics of sentiment: we identify the events when the target companies experience extreme changes in sentiment, and find that these events are often correlated with movements of sentiment in the aforementioned groups of related companies. To analyse the market movements associated with sentiment, we further acquire daily Bloomberg market data on the target companies during the same period. We relate the sentiment events to market movements, and find that these events in expectation induce unusual movements in return and volatility in not only the companies experiencing the events themselves, but also at the sectorial or group level. In light of those findings, we would also like to advocate the pipeline of analysis used in this paper as a general-purpose, real-time framework for network analysis of financial news sentiment correlation and the corresponding market movements. 

\section*{Results}

\subsection*{Evolution of Sentiment}
\begin{figure}[h!t]
  \subfloat[]{\includegraphics[width=1\linewidth]{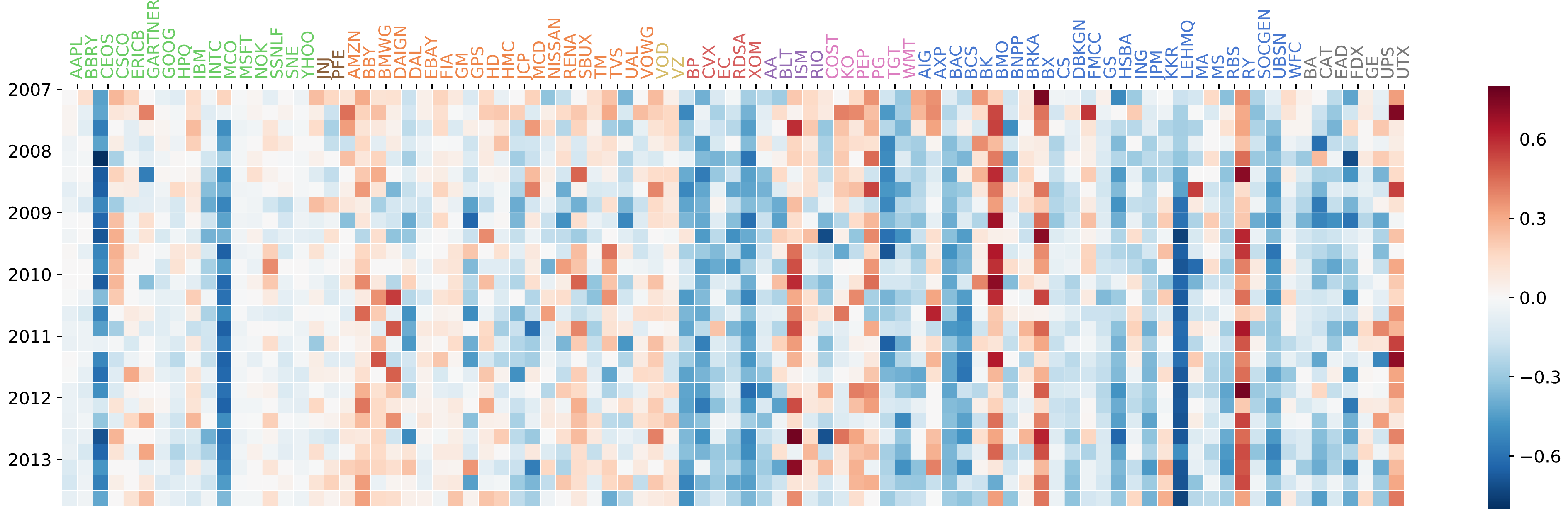} }\\
  \subfloat[]{\includegraphics[width = 0.45\textwidth]{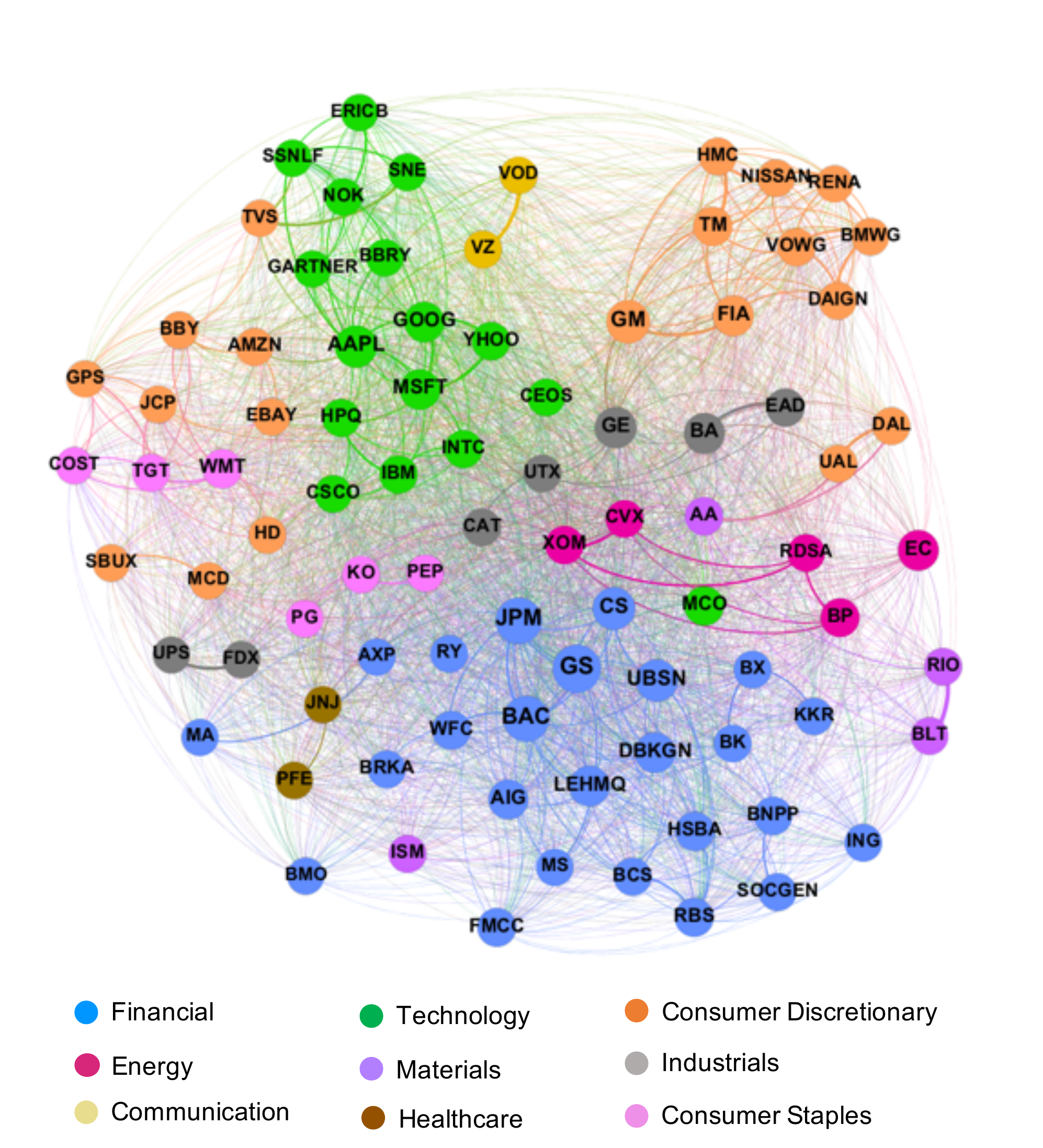}}
  \subfloat[]{\includegraphics[width = 0.45\textwidth]{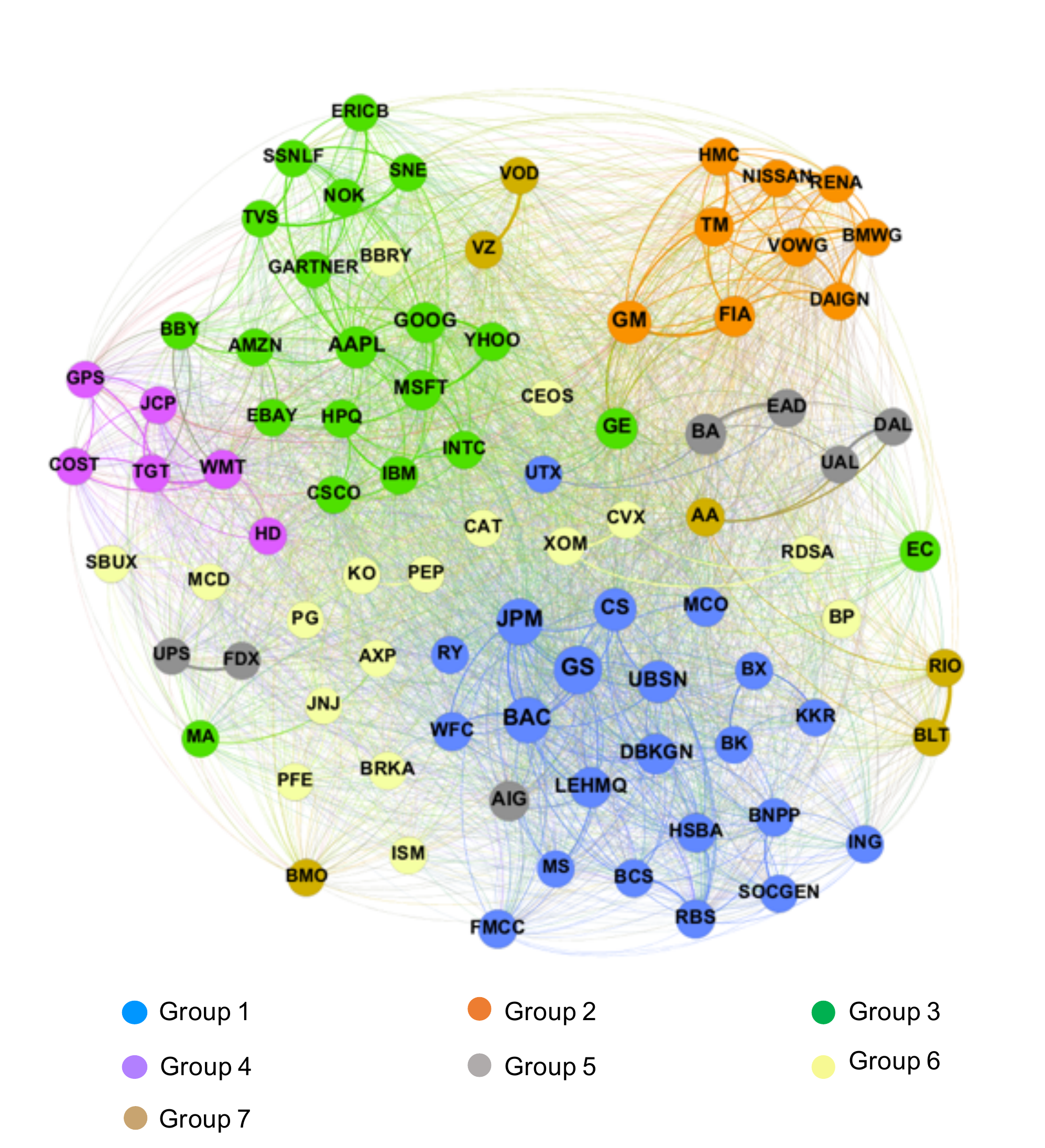}}
  \caption{News co-occurrence network and sentiment distribution. (a) Sentiment distribution for 87 target companies in 27 quarters from 2007 to 2013. Colors in x-tick represent different sectors; (b) News co-occurrence network colored with sector; (c) News co-occurrence network colored with detected groups.}
  \label{fig:general_figure}
\end{figure}

Through NLP techniques, we detect entities (i.e., companies) from financial news articles on Reuters, and merge entities that co-reference the same company. We compute a sentiment value for each of the companies that appears in a news article (see Methods for details of the NLP processing). We end up selecting 87 companies in 9 sectors, listed in Table \ref{tab:long}. In Figure \ref{fig:general_figure}(a), we visualize the evolution of news sentiment of the 87 companies for 27 quarters in our data. Each cell represents the average sentiment of the corresponding company in a specific quarter. Sentiment value is represented using cell color, with positive sentiment in red and negative sentiment in blue. 
We observe that banks in the U.S. (\textit{LEHMQ, JPM, GS}) experienced strong negative sentiment around the financial crisis in September 2008, which indicates that those US banks were significantly affected in the crisis. However, the sentiment of Canadian banks (\textit{BMO, RY}) did not suffer as much from the financial crisis, which is in line with the perception at the time \cite{kevinwebsite}. 

\begin{figure}[t] 
  \subfloat[]{\includegraphics[width =0.3\textwidth]{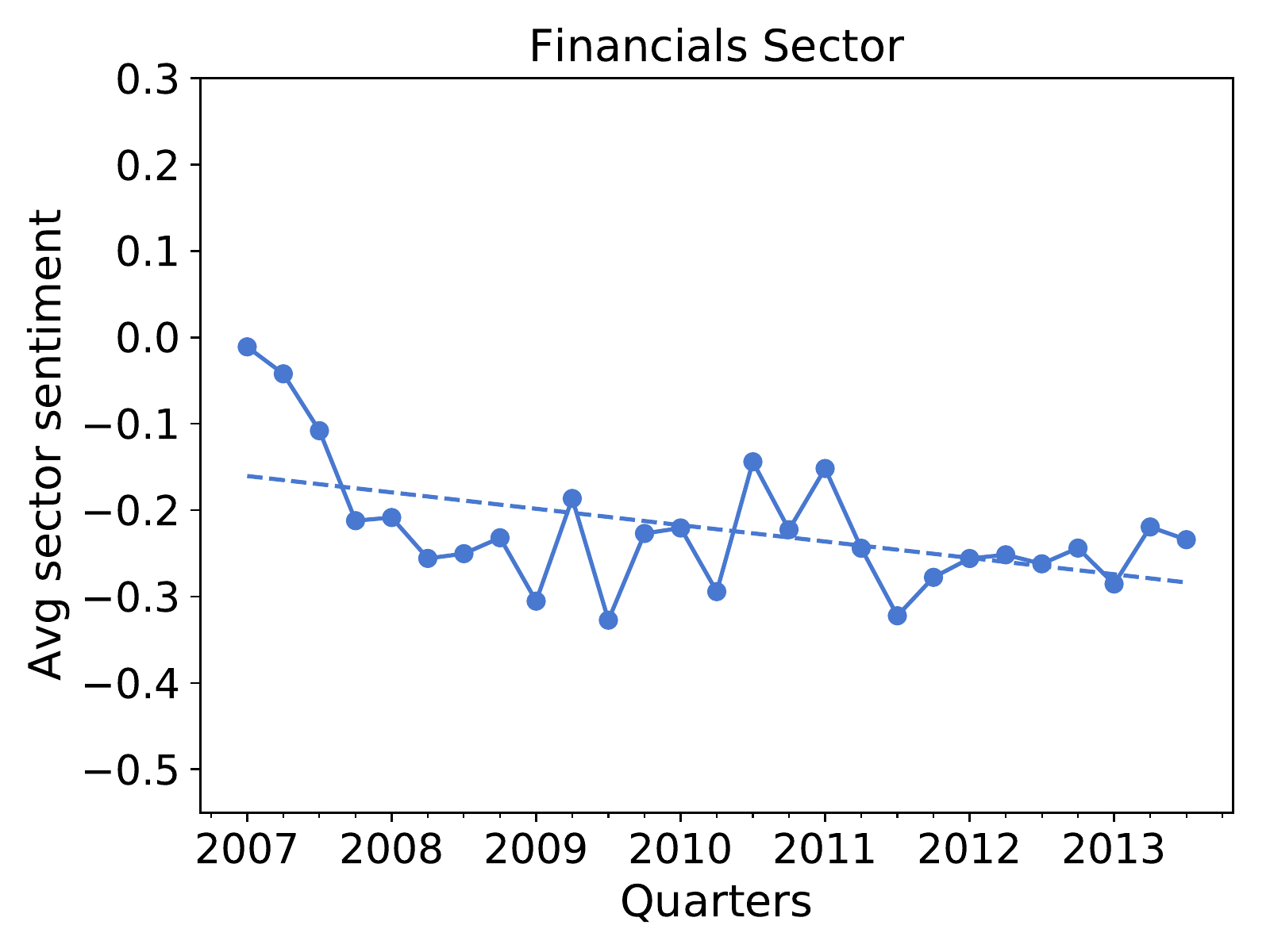}}
  \subfloat[]{\includegraphics[width =0.3\textwidth]{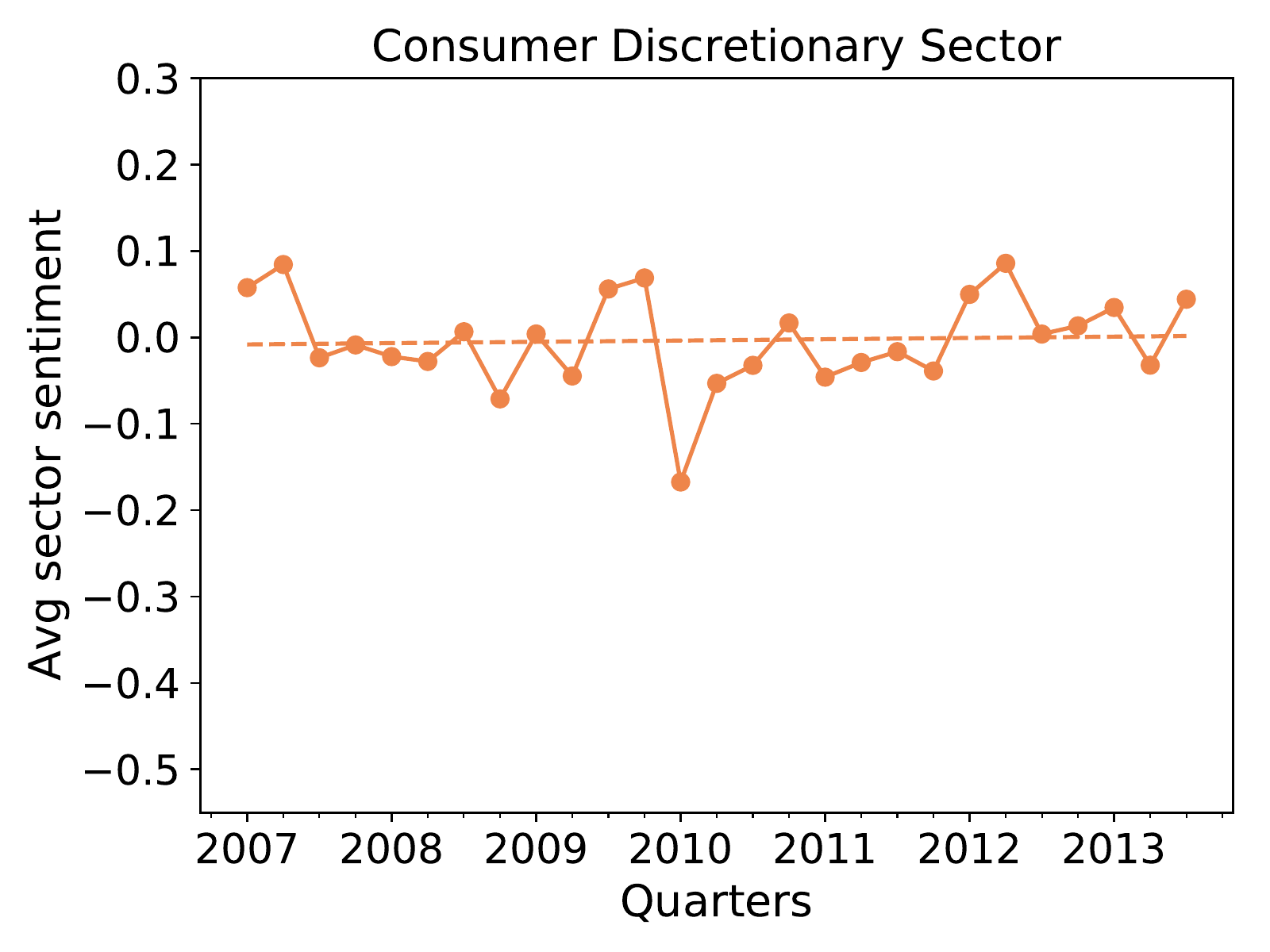}}
  \subfloat[]{\includegraphics[width =0.3\textwidth]{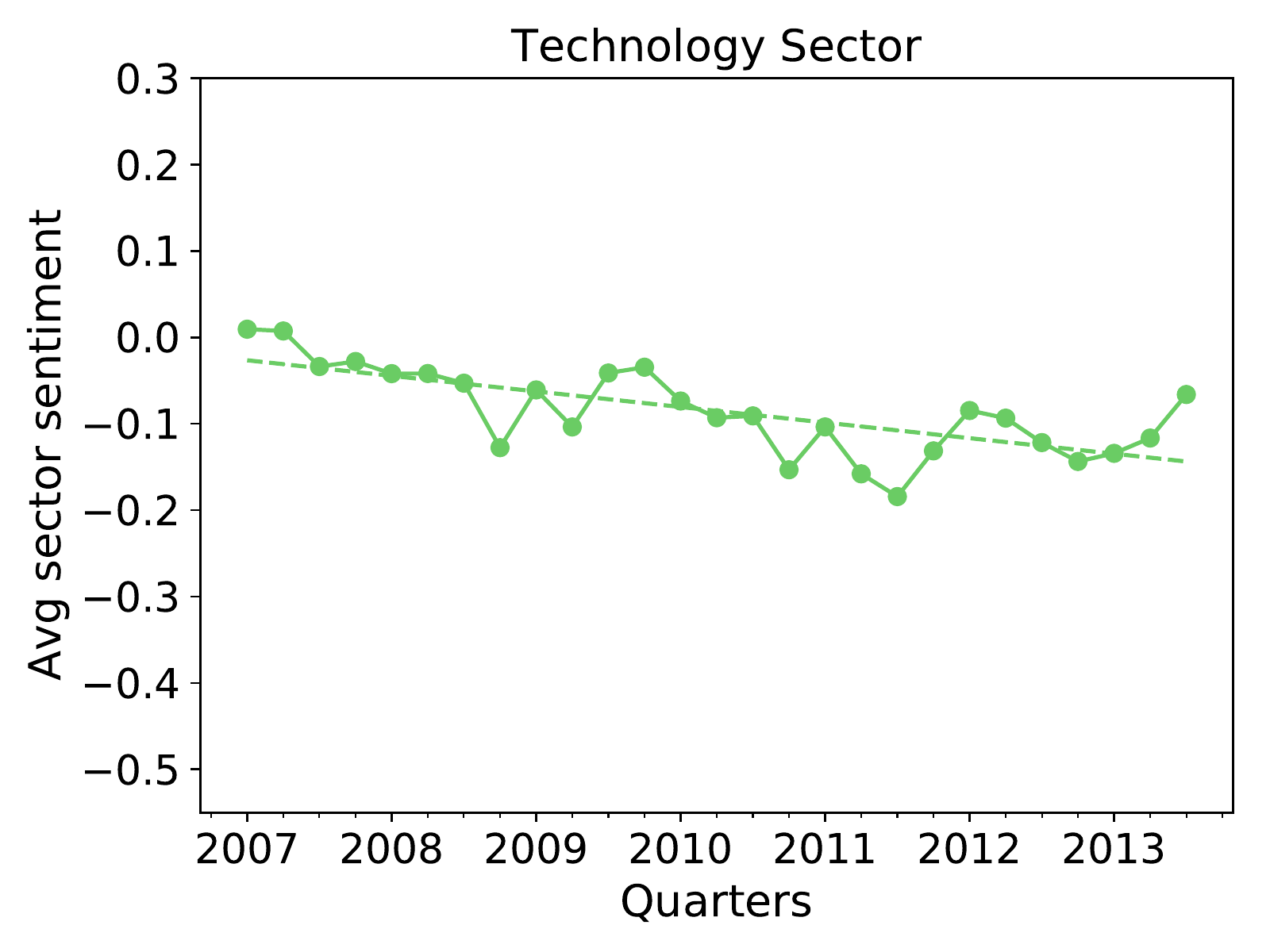}}\\
  \subfloat[]{\includegraphics[width =0.30\textwidth]{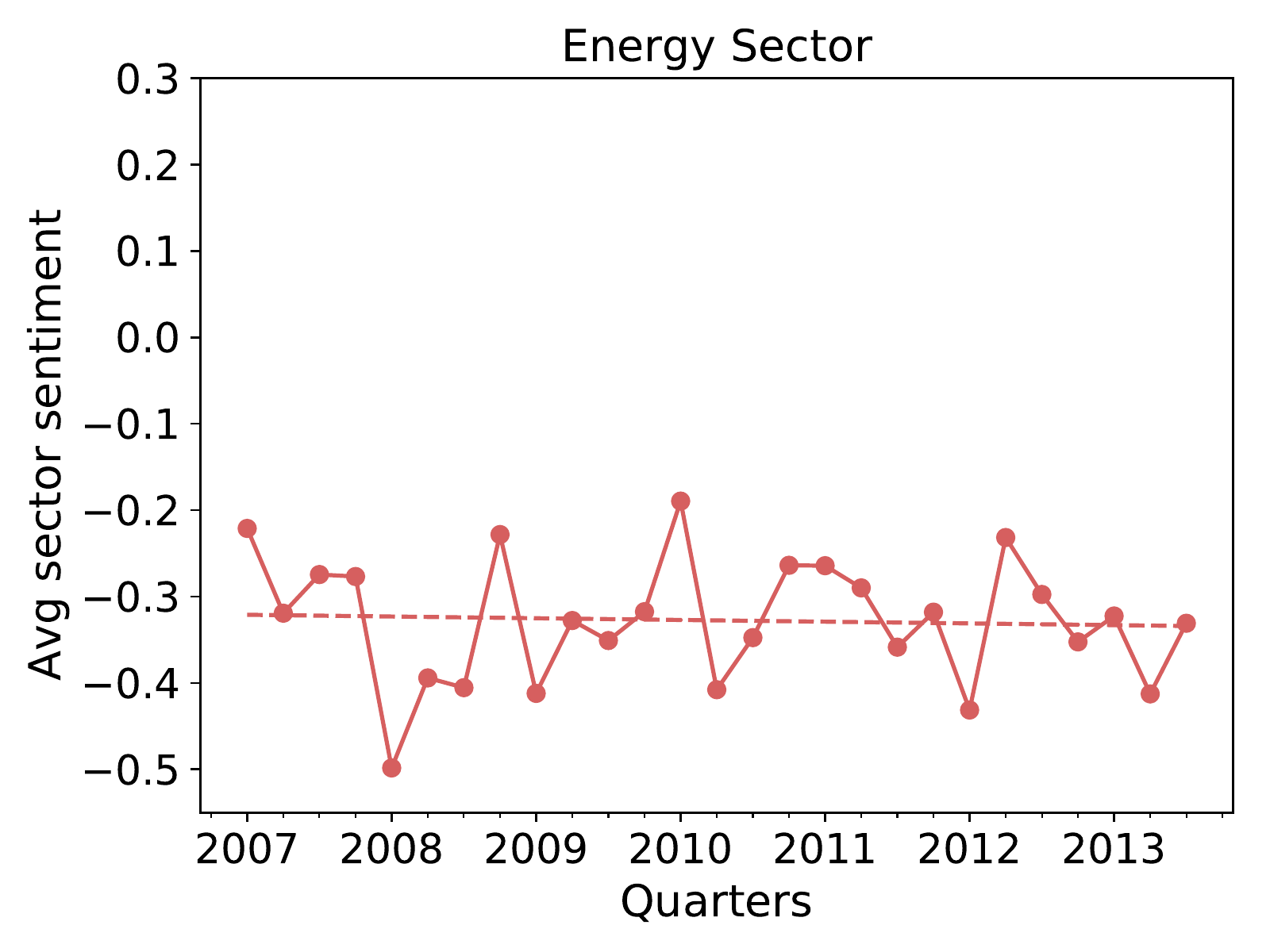}}
  \subfloat[]{\includegraphics[width =0.30\textwidth]{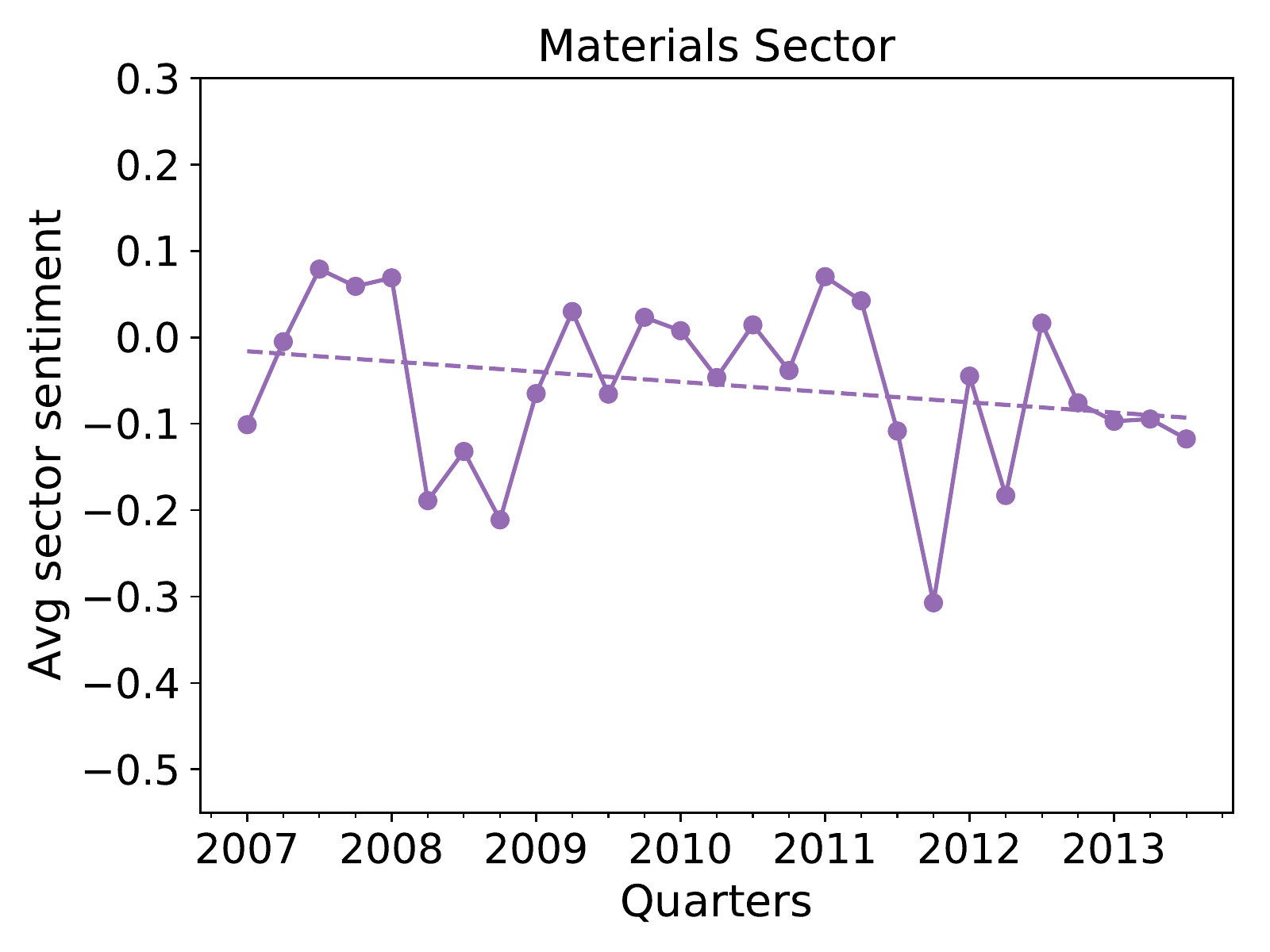}}
  \subfloat[]{\includegraphics[width =0.30\textwidth]{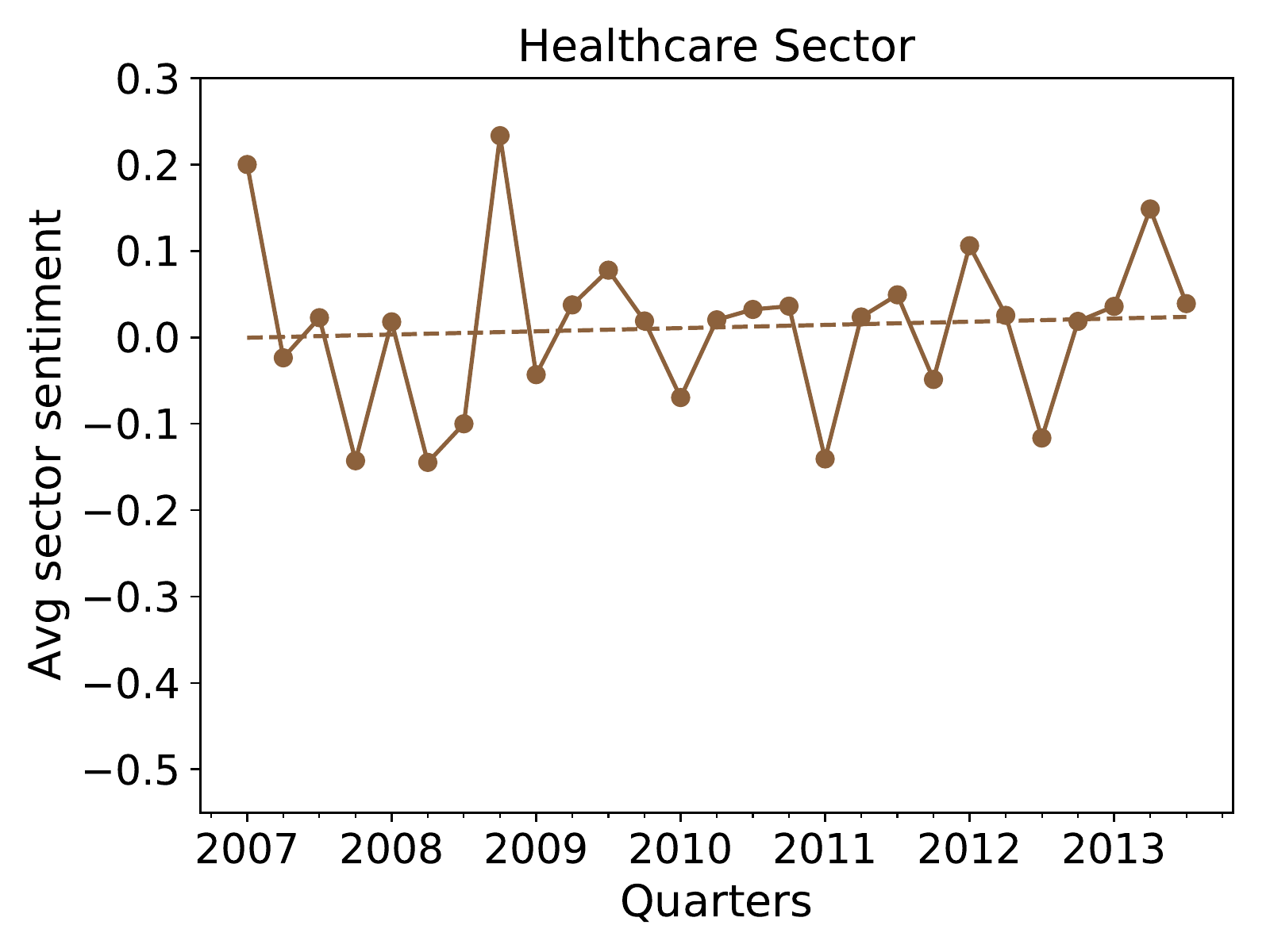}}\\
   \subfloat[]{\includegraphics[width =0.30\textwidth]{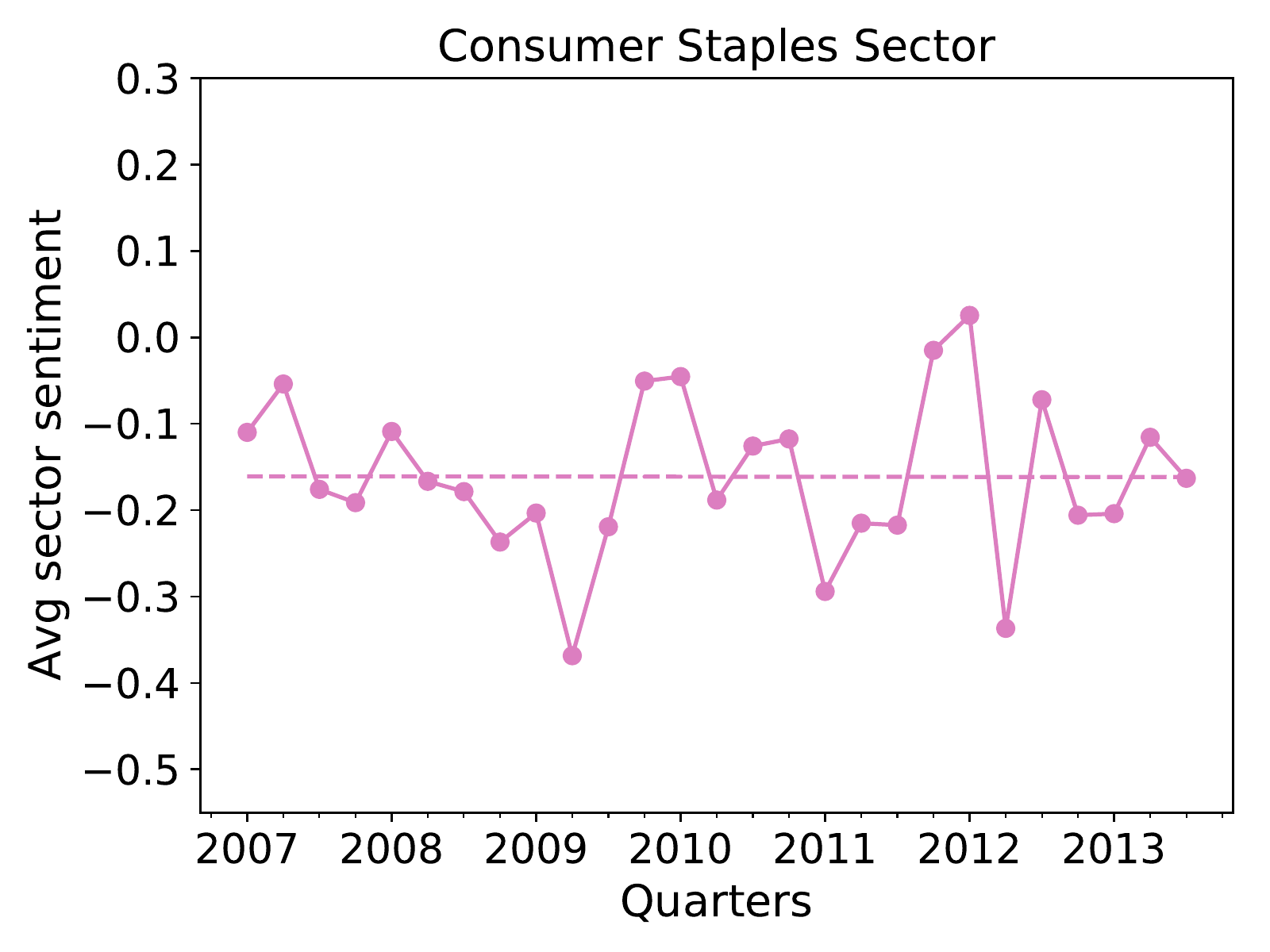}}
  \subfloat[]{\includegraphics[width =0.30\textwidth]{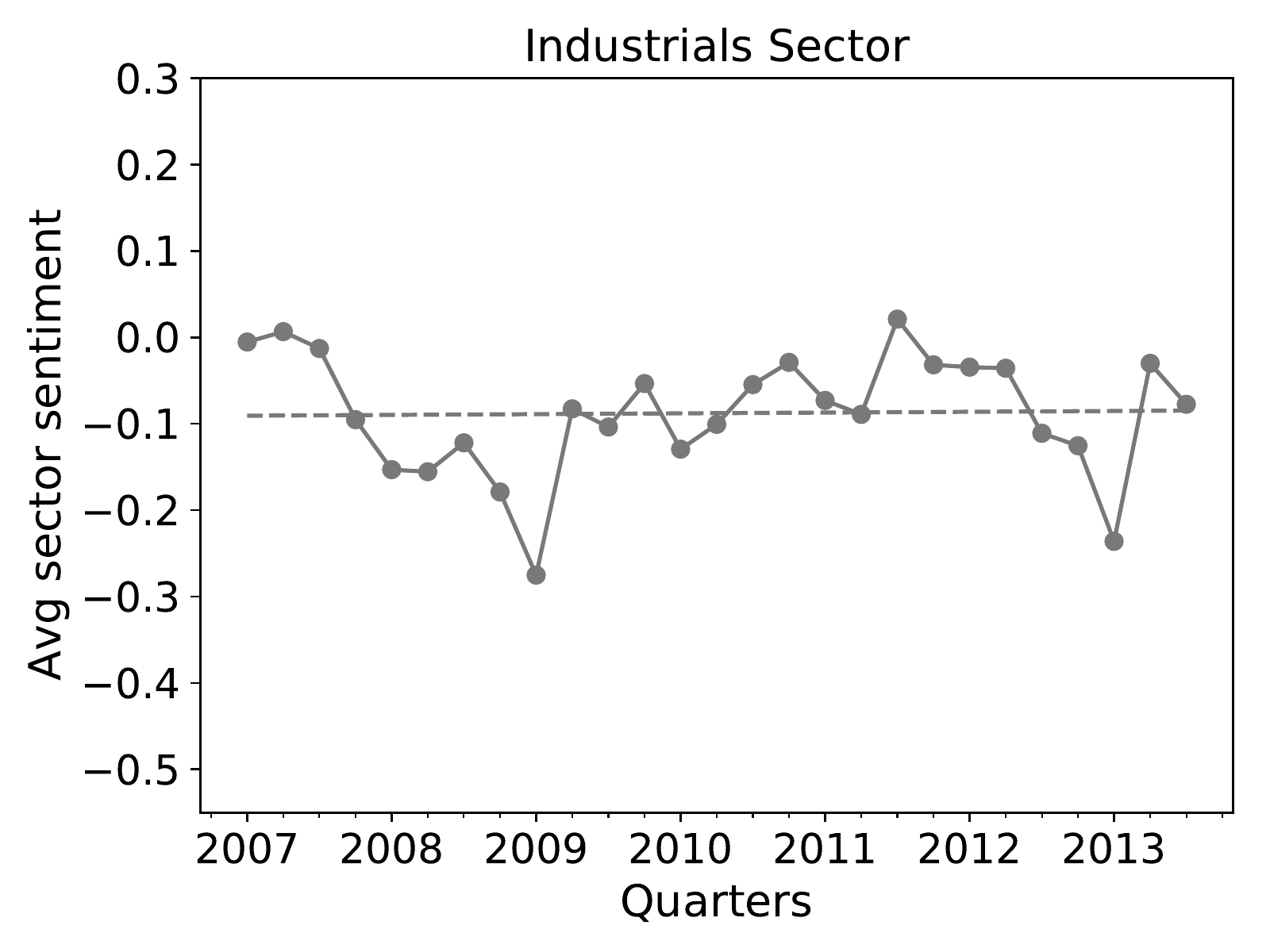}}
  \subfloat[]{\includegraphics[width =0.30\textwidth]{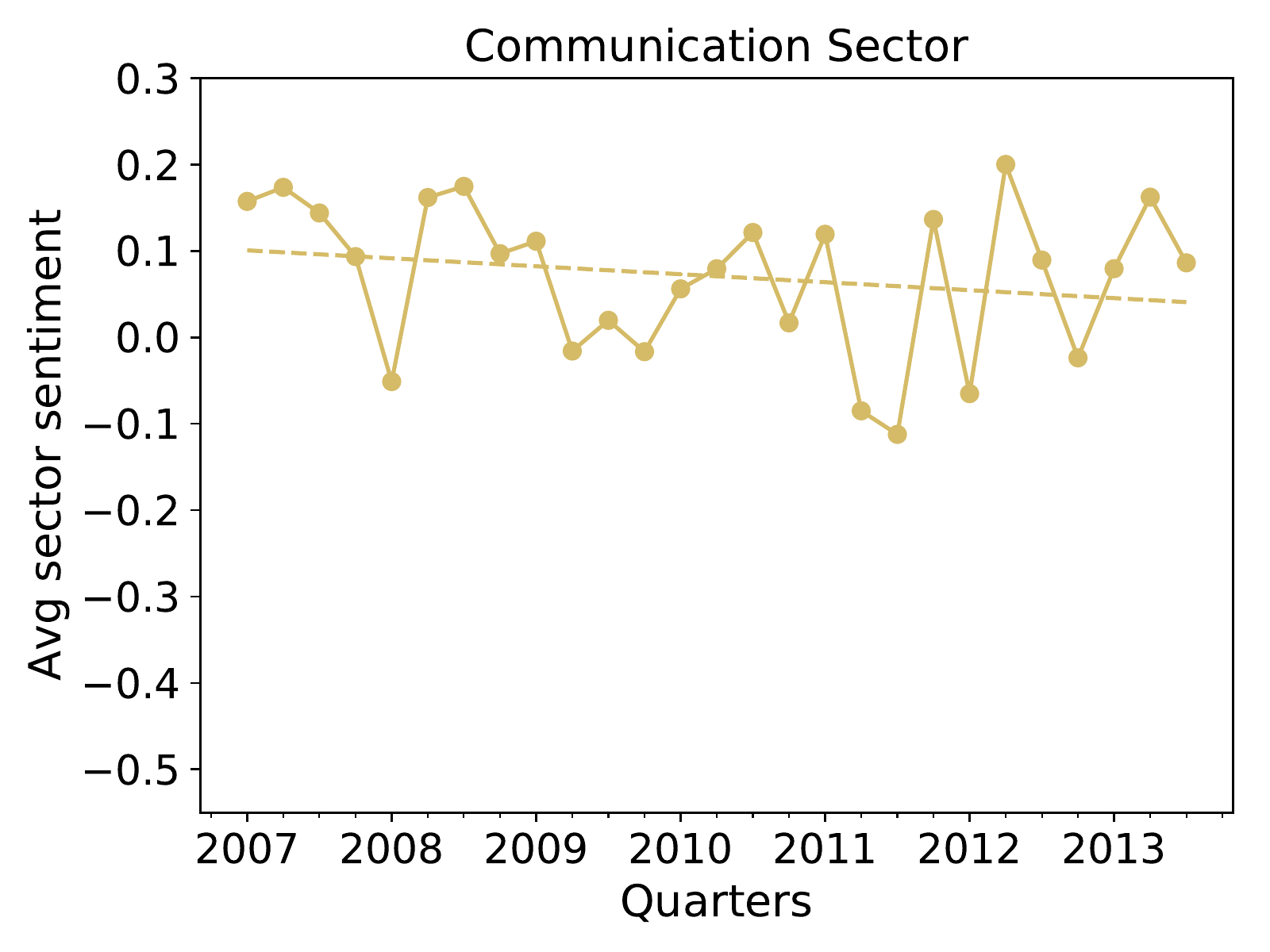}}
  \caption{Evolution of average sentiment for the 9 sectors.}
  \label{fig:sector_sentiment}
\end{figure}

In Figure \ref{fig:sector_sentiment}, we furthermore present the average sentiment of each sector for each quarter, obtained by averaging the sentiment of all the companies in the sector.
We observe that, companies in the ``Technology'' sector had a relatively stable sentiment distribution along all the quarters, but those in the ``Financial Services'' sector was more volatile. 
Overall, most sectors had negative sentiment in the financial crisis quarters, which clearly implied the impact of the crisis. Sentiment of the ``Energy'' and ``Financial Services'' sectors were negative in most quarters, which is consistent with sentiment experienced by individual companies in those sectors.

\subsection*{News Co-occurrence Network}
The financial market is a system with complex inter-dependencies between companies. Co-appearance of companies in the news may reveal the underlying relations between them, thus provide a unique perspective to study such inter-dependencies. To this end, we first construct a news coverage matrix where each row represents a company and each column represents a news in the period of study. The $ij$-th entry of the matrix is the number of news articles in which company $i$ appear on news $j$.
From the news coverage matrix, we construct a weighted news co-occurrence network using the first year (2007) of the news data, where nodes represent companies and edges capture the relationships between pairs of companies. The edge weight $e_{ij}$ between a pair of nodes $i$ and $j$ is defined as the cosine distance between the corresponding row vectors of the news coverage matrix. 

To analyse the structure of the news co-occurrence network, we apply community detection \cite{blondel2009fast} on the network (we detail the technique in Methods) to yield 7 major groups of companies, a number that is comparable to the 9 ground-truth sectors: it is worth noting that some sectors (such as Communications and Healthcare. See Table \ref{tab:long}) are very small, and as such forcing the community detection method to produce the exact number of corresponding groups by tweaking the timescale parameter leads to unsatisfactory partitioning of groups. We present the groups obtained in Figure \ref{fig:general_figure}(c), and we also present the visualisation of the companies' ``ground-truth'' Bloomberg sector grouping in Figure \ref{fig:general_figure}(b). We use the software package Gephi \cite{ICWSM09154} for the visualisations, and companies in the same sector or group are labeled with the same color. 

Through the comparison between Figure \ref{fig:general_figure}(b) and Figure \ref{fig:general_figure}(c), we observe a high-level consistency between the major company groups detected in the co-occurrence network and the ``ground-truth'' sectors (see Figure \ref{fig:sector_cluster} in Supplementary information for a detailed distribution of sectors in each company group). Furthermore, we also find that while the median edge weight between two companies not belonging to the same sector (\textit{out-sector}) is $0.00229$, the median weight between two companies belonging to the same sector (\textit{in-sector}) is $0.0157$ (see Figure \ref{fig:weightdistr} in Supplementary Information for the probability distributions of weights). These suggest that we may largely recover the sectors simply by looking at how often the companies are mentioned together within a rather short timespan of a year.

Nonetheless, the network reveals more interesting information when the groups and sectors \textit{differ}, which potentially contains insights beyond what sector classification offers. Specifically, taking explicitly into account of the different weight distributions for in-sector and out-sector companies (that the in-sector company pairs are expected to have larger weights), we use statistical criteria to isolate the outlier company pairs of the weight distributions. These outliers are either the company pairs \textit{not} belonging to the same sector yet exhibiting extraordinarily strong co-occurrence (for the case of outliers in the out-sector distribution) or company pairs belonging to the same sector but exhibiting particularly strong links, which potentially reveals more fine-grained information about the relations of companies within sectors (for the case of outliers in the in-sector distributions) and we include the details of the filtering procedure in Methods. We show the lists of interesting company pairs we find in Tables \ref{tab:outsectoroutliers} and \ref{tab:insectoroutliers} of Supplementary Information. To give some specific examples, in the former category (Table \ref{tab:outsectoroutliers}), we find that while being classified as ``Consumer Discretionary'', AMZN (Amazon) has strong links with both Technology companies (e.g. GOOG (Google) and AAPL (Apple)) and retail companies (e.g. WMT (Walmart)), which are nominally classified under ``Consumer Staples''; this actually is reflective of the market position and operating model of AMZN. We also find interesting links such as BKRA  (Berkshire Hathaway) - PG (Proctor \& Gamble) and BKRA - KO (Coca-Cola), which could be explained by the large stake BKRA holds for both PG and KO. In the latter category (Table \ref{tab:insectoroutliers}), we find that especially strong links are often between companies that are often direct competitors and/or with considerable overlap in business interests within the same sector, e.g., VOD (Vodafone) - VZ (Verizon), and GS (Goldman Sachs) - MS (Morgan Stanley).

We also directly examine the composition of the \textit{clusters} and compare and contrast with the ground-truth classification in Table \ref{tab:clusterfirstyear} and Figure \ref{fig:sector_cluster} in the Supplementary information: Groups 1 and 2 are dominated by Financials and Consumer Discretionary companies (actually Group 2 is dominated by automobile manufacturers, which are a sub-division within the broader Consumer Discretionary sector. This suggests that finer-grained information can be obtained from the co-occurrence network). On the other hand, Groups 3-5 are dominated by a few different sectors, where such differences are usually interpretable: for example, while classified by Bloomberg in the ``Consumer Discretionary'' category, eBay (EBAY) and Amazon (AMZN) are grouped with other companies in ``Technology''. Similarly, we also observe that Group 4 is dominated by more traditional offline retail companies, even though their sectors are split between ``Consumer Discretionary'' and ``Consumer Staples''. Finally, we see a larger amount of disagreement in the final two groups (Groups 6 and 7), where many clusters of companies too small to be standalone groups are included. The discussions here highlight that the news co-occurrence network may provide additional insight into the relations between the companies that is not reflected in the sector segmentation; more importantly, the network can be constructed dynamically, and is thus capable of capturing the evolution of the relations between companies over time.

This latter point motivates us to build a co-occurrence network for each quarter of the year and investigate the evolution of positions of companies in the network using the eigenvector and betweenness centrality measures. In Figure \ref{fig:centrality_single} in Supplementary information, we show results for six top companies from Technology and Financial sectors (Google, Microsoft, Apple, Goldman Sachs, JP Morgan, and Lehman Brothers). It can be seen that Google and Apple had an increasing centrality in the network while Microsoft had a decreasing one in the seven years investigated. The three financial companies had decreasing centrality and Lehman Brothers faded quickly after its bankruptcy during the crisis. This therefore provides a unique angle to investigate the dynamic standings of companies in the financial markets.


We further study the structure of the news co-occurrence network over time, and the results are summarised in Figure \ref{fig:quarter_network} in Supplementary information. Figure \ref{fig:quarter_network}(a) shows the average degree of the co-occurrence network for each quarter. The network average degree reached the peak in the last quarter of 2008, which corresponds to the escalation of the financial crisis. This might be due to more companies being reported and discussed together during events of catastrophic failure in the market \cite{almog2019structural}. The increasing connectedness of the network can also be seen from its clustering coefficient and average path length shown in Figure \ref{fig:quarter_network}(b).
In terms of clustering structure, we compare in Figure \ref{fig:quarter_network}(c) the auto-detected groups based on the co-occurrence network for different quarters against the ground-truth sectors, using normalised mutual information (NMI) and F1 as measures of group similarity. We observe that the similarity is generally increasing. This might be contributed by the general increasing number of financial news articles over the period of the seven years, which helps reflect more accurately the relations between the companies.

\subsection*{Sentiment Dynamics in Related Companies}
The inter-dependencies within the financial market may suggest that a strong media sentiment towards one company, often caused by a significant real-world event, could have a spillover effect on related companies. To this end, we first detect days on which there is a significant upward or downward change in the sentiment score of a target company, which we denote as \textit{event days}. There are a few key rationales why we focus on sentiment \textit{events} instead of the entire day-by-day sentiment time series: firstly, by focusing only on the days when there are significant changes in sentiment, we implicitly filter out the noises that are inevitably present in the sentiment time series. Secondly, sentiment time series especially between related companies can be themselves correlated to some extent, as financial news outlets such as Reuters often simultaneously report a number of related companies. In this paper we are primarily interested in investigating how strong sentiment in one company spills over to related companies, hence such effect would be a confounding factor. However, we find that by focusing on the sentiment events only, we largely remove the said correlation as we find the sentiment \textit{events} of different companies are largely uncorrelated (See Figure \ref{fig:sentimenteventcorr} in Supplementary Information for the pairwise correlation matrix of all target companies, and the readers are referred to Methods section for details).

We then examine the average sentiment of related companies before and after the event days, where related companies are defined as nearest neighbours of the target company in the news co-occurrence network (see Methods section for more details about the event days and related companies). It is worth noting that the related companies do \textit{not} need to experience sentiment events themselves. 
Finally, for ease of presentation, we aggregate results for different target companies by the groups shown in \ref{fig:general_figure}(c) and Table \ref{tab:clusterfirstyear} in Supplementary information.

\begin{figure}[t]
  \subfloat[]{\includegraphics[width = 0.22\textwidth]{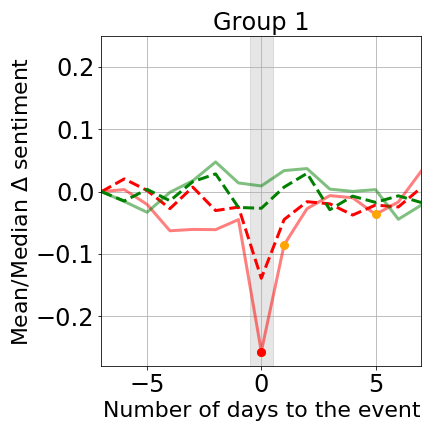}}
  \subfloat[]{\includegraphics[width = 0.22\textwidth]{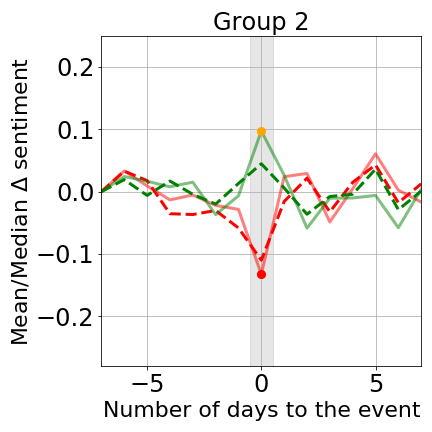}}%
\subfloat[]{\includegraphics[width = 0.22\textwidth]{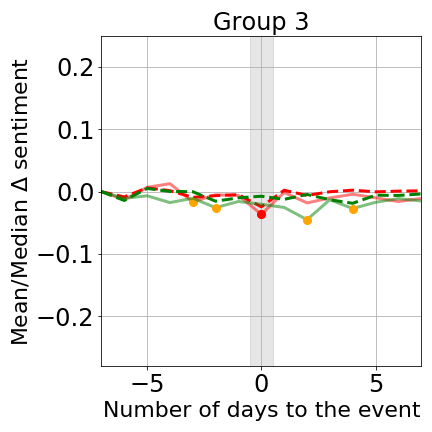}}%
    \subfloat[]{\includegraphics[width = 0.22\textwidth]{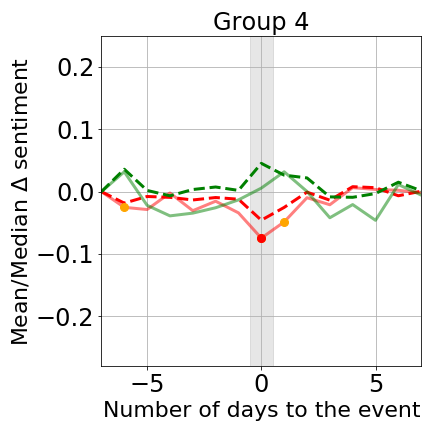}}%

  \subfloat[]{\includegraphics[width = 0.22\textwidth]{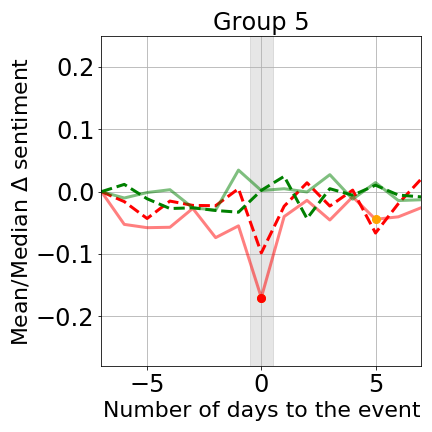}}
  \subfloat[]{\includegraphics[width = 0.22\textwidth]{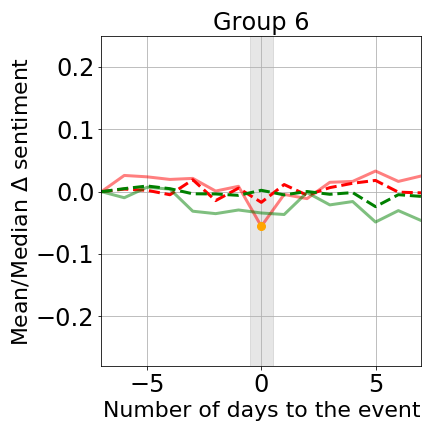}}%
\subfloat[]{\includegraphics[width = 0.22\textwidth]{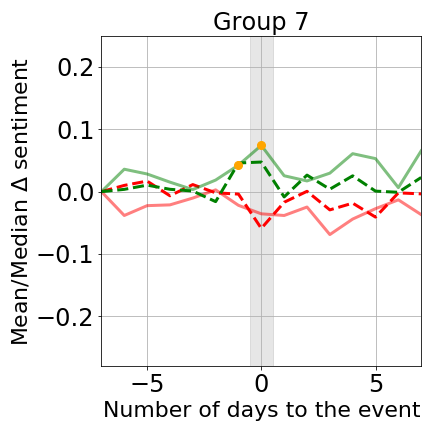}}%

  \includegraphics[width = 0.8\textwidth]{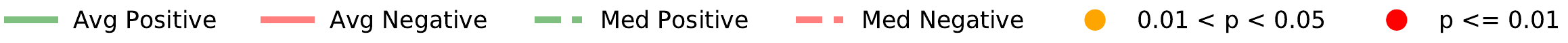}

  \caption{Change in mean and median group aggregate sentiment score of the major groups.}
  \label{fig:sentimentdiffusion}
\end{figure}

The results are presented in Figure \ref{fig:sentimentdiffusion}. We conduct a non-parametric Mann-Whitney U-test\cite{mann1947test} on the null hypothesis that the change in group aggregate sentiment is zero, and marked statistically significant points in red ($p < 0.01$) and orange ($ 0.01 \leq p < 0.05$). We find that there exists a clear positive relationship between the sentiment of the individual company and the group sentiment for several company groups, although the strength of this relationship varies across groups. For example, in Group 1 which mainly consists of financial services companies, negative sentiment change in one company often correlates with large movement across the group, while positive sentiment does not correlate with statistically significant group movement. The same patterns are observed for companies in Groups 3, 5 and 6. On the other hand, for Groups 2 and 4 which are both related to the consumer product-related companies (both Consumer Discretionary and Consumer Staples sectors - See Figure \ref{fig:sector_cluster} in Supplementary information), the synchronised group sentiment movements from positive and negative events of individual companies are often more symmetrically felt. 

\floatsetup[figure]{subcapbesideposition=top}

\floatsetup[figure]{subcapbesideposition=top}
\begin{figure}[h!btp]
  \subfloat[]{\includegraphics[width = 0.22\textwidth]{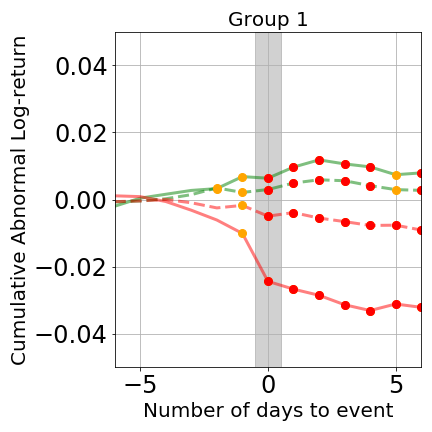}}
  \subfloat[]{\includegraphics[width = 0.22\textwidth]{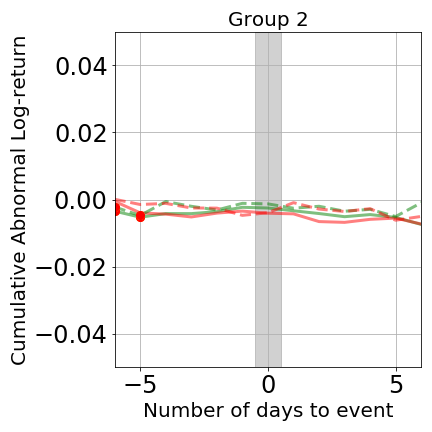}}
  \subfloat[]{\includegraphics[width = 0.22\textwidth]{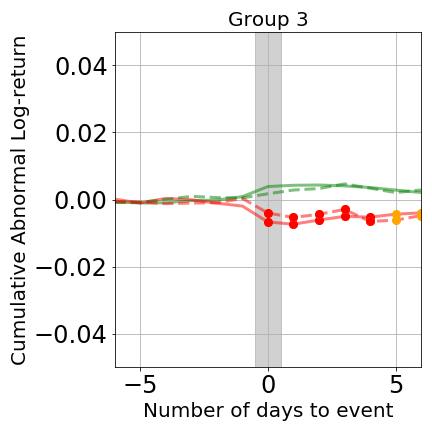}}
  \subfloat[]{\includegraphics[width = 0.22\textwidth]{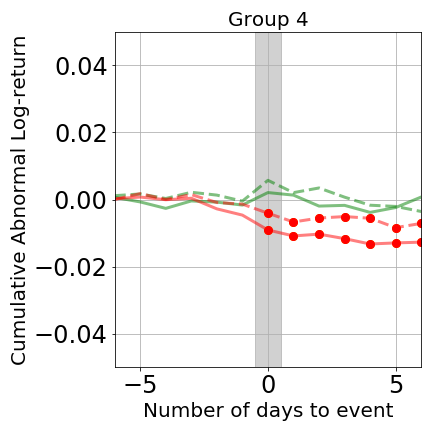}}

  \subfloat[]{\includegraphics[width = 0.22\textwidth]{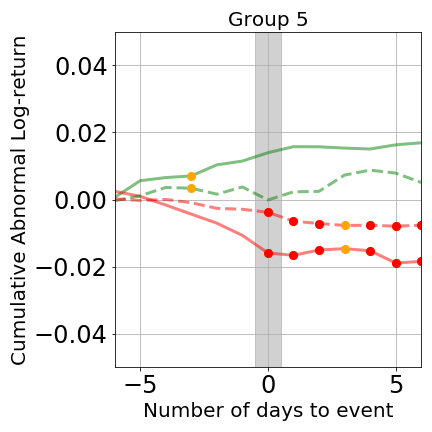}}
  \subfloat[]{\includegraphics[width = 0.22\textwidth]{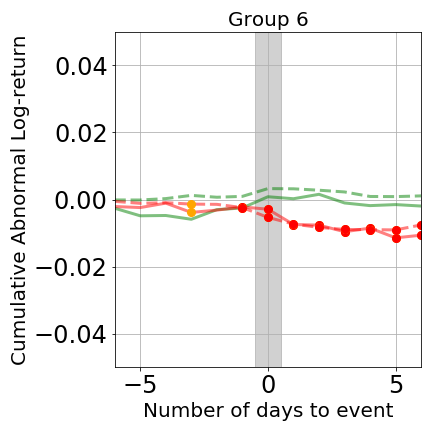}}
  \subfloat[]{\includegraphics[width = 0.22\textwidth]{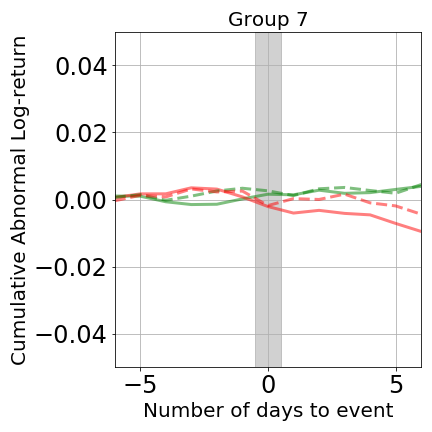}}
  \caption{Change in mean and median \textbf{CAR} of the companies experiencing sentiment events, computed over all events over the time period and averaged over each company group.}
  
    \includegraphics[width = 0.8\textwidth]{newfigures/legend.pdf}

  \label{fig:individualmarketimpact}
\end{figure}

\subsection*{Market Movements Associated with Sentiment}

Media sentiment is often associated with potential market movements. To begin with, it is important to investigate the sources of the media sentiment, for sentiment would be of limited usefulness to either explain or potentially predict market performance, if a significant amount of the sentiment is itself derived from commenting the market (especially so since Reuters does publish articles purely commenting on market performance). We argue that at least from a typical sample, most sentiment,  \textit{at least on event days where there are massive changes in sentiment}, is primarily derived from underlying fundamental events, instead of trivially mirroring the market performance (see Methods section for full methodology of the experiments conducted and conclusions drawn). 

\begin{figure}[t]
  \subfloat[]{\includegraphics[width = 0.22\textwidth]{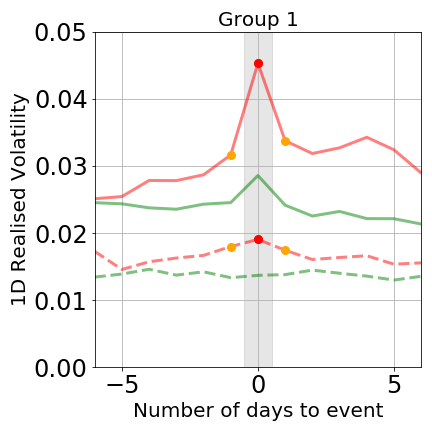}}
  \subfloat[]{\includegraphics[width = 0.22\textwidth]{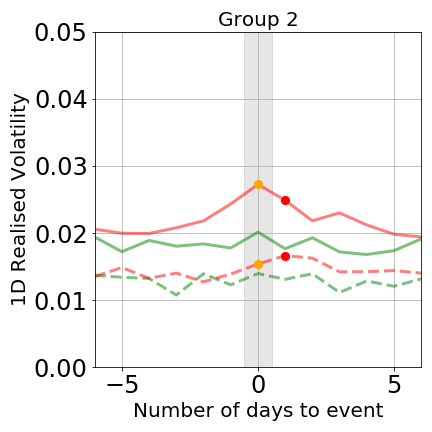}}
  \subfloat[]{\includegraphics[width = 0.22\textwidth]{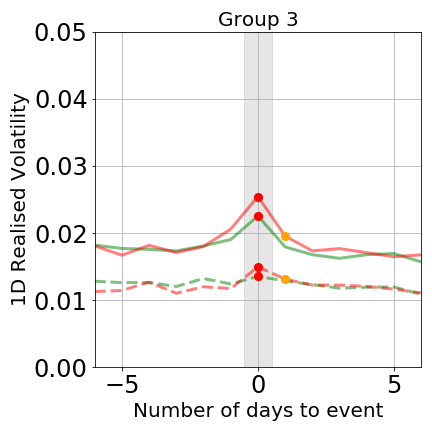}}
  \subfloat[]{\includegraphics[width = 0.22\textwidth]{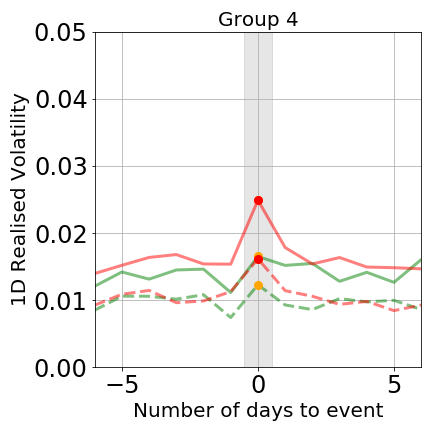}}

  \subfloat[]{\includegraphics[width = 0.22\textwidth]{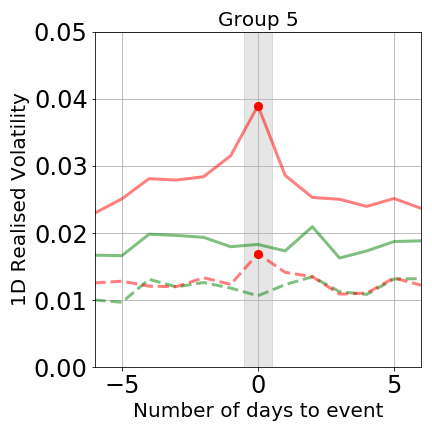}}
  \subfloat[]{\includegraphics[width = 0.22\textwidth]{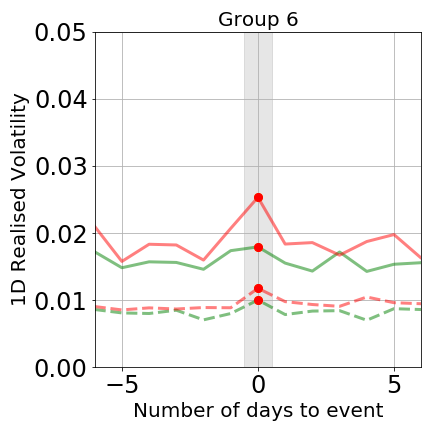}}
  \subfloat[]{\includegraphics[width = 0.22\textwidth]{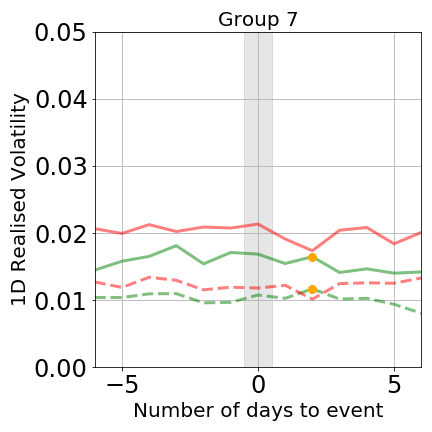}}
  
  \includegraphics[width = 0.8\textwidth]{newfigures/legend.pdf}
  \caption{Change in mean and median \textbf{daily volatility} of the companies experiencing sentiment events, computed over all events over the time period and averaged over each company group.} 
  \label{fig:individualmarketimpactvol}
\end{figure}

We then compute the cumulative abnormal return (CAR) and daily realised volatility for the individual companies around the days of sentiment events 
and aggregate the results based on the 7 company groups mentioned in the previous section (see Methods for details). 
The results are shown in Figure \ref{fig:individualmarketimpact} for CAR and in Figure \ref{fig:individualmarketimpactvol} for daily volatility. Following the methodology of Ranco \textit{et al.}\cite{ranco2015effects}, we test statistical significance on CAR. However, differing from the z-test used in their studies which entails Gaussianity assumption, we again use the Mann-Whitney U-test\cite{mann1947test} for statistical significance, both in conformity with the previous section and to avoid making any assumption on the underlying distributions. In Figure \ref{fig:individualmarketimpact}, we test against the null hypothesis that the CAR is zero; In Figure \ref{fig:individualmarketimpactvol}, we make the assumption that the volatility before the event days are drawn from a stationary distribution, and test against the hypothesis that the volatility after the events are drawn from the same distribution. Statistically significant points are marked in red ($p < 0.01$) and orange ($ 0.01 \leq p < 0.05$).

The results demonstrate that for most groups, there exists a clear relation between media sentiment and market performance: while there are few statistically significant deviations from zero in CAR before sentiment events (with a few exceptions), the CAR patterns diverge remarkably after the sentiment events. Generally speaking, positive sentiment promotes upward price movements whereas negative one promotes price declines and as a result both types of events elicit elevation in volatility. For some company groups, this effect lingers on for a few days as the CAR continues to decline or improve gradually after the initial change on the event day. This effect again differs across different groups: it is, for example, most prominent in Finance-dominated Group 1, which seems to suggest that financial companies are more sensitive to sentiment change, both in terms of the correlated movement in group sentiment (as shown in Figure \ref{fig:sentimentdiffusion}(a)) and the market reaction (as shown in Figure \ref{fig:individualmarketimpact}(a) and Figure \ref{fig:individualmarketimpactvol}(a)). Similar to Figure \ref{fig:sentimentdiffusion} where negative events are shown to be associated with greater movement in group sentiment, we also observe that negative events also correlate with larger market movements. It is also interesting to observe that in a number of cases the perceived effect of sentiment on volatility is not only observed \textit{on} the event day, but \textit{after} event days as well, suggesting the lingering effect of extreme sentiment movements and potential predictive uses of large sentiment movements for market participants.

\floatsetup[figure]{subcapbesideposition=top}

The correlation of media sentiment across companies suggests that market movements may also be studied at a collective level for groups of companies.
At the collective level, the complex interactions between companies preclude the simple analysis of average statistics as in the individual case. 
Instead, we computed the implied probability density function (PDF) of the histograms of abnormal return (AR) and volatility in each group on the days before, on, and after the days when there are extraordinary changes in group aggregate sentiment events (see Methods for detail).
We discuss in this section results for one company group, i.e., Group 2, and results for several other groups are presented in Figures \ref{fig:pxgrouppositive}-\ref{fig:volgroupnegative} in Supplementary information.

From Figure \ref{fig:group2dist}, we note that while the effect at the group level is unsurprisingly weaker, there are still noticeable deviations in the profiles of distributions on event days. This suggests that links exist not only between an individual company's sentiment and price movement, but also at a collective level. Specifically, for the Industrial-dominated Group 2, we observe that when the group-aggregated sentiment experiences extraordinary changes in one direction, there is a noticeable shift in the probability mass of the distribution of AR towards the opposite direction. For volatility, there is a noticeable elevation in probability mass on both types of sentiment events, suggesting that on these days the market movements of the group as a whole are likely to be more volatile. Such observations, especially the elevation in volatility, are also observed in a number of other groups, as shown in Figures \ref{fig:pxgrouppositive}-\ref{fig:volgroupnegative}. Interestingly, on Figure \ref{fig:individualmarketimpact}, Group 2 seems to exhibit the least correlation between sentiment and market movement at the individual company level;
this suggests that analysis without considering network effects might prove inadequate in our context.

\floatsetup[figure]{subcapbesideposition=top}
\begin{figure}[t]
  \subfloat[]{\includegraphics[width = 0.20\textwidth]{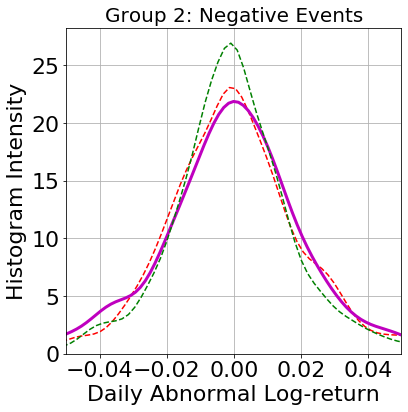}}
  \subfloat[]{\includegraphics[width = 0.20\textwidth]{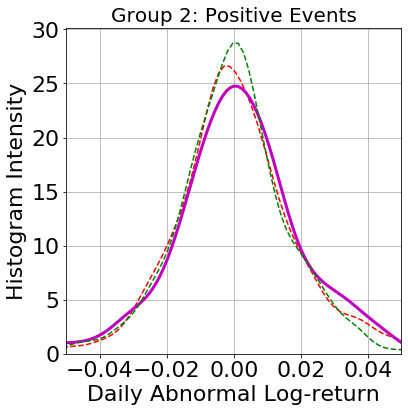}}
  \subfloat[]{\includegraphics[width = 0.20\textwidth]{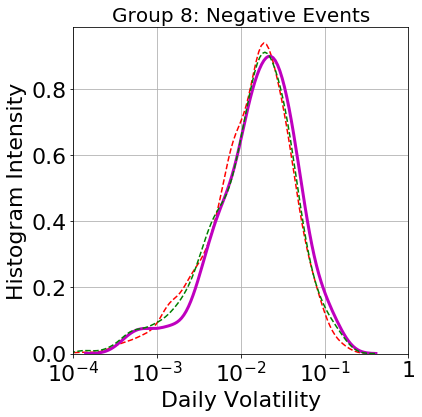}}
  \subfloat[]{\includegraphics[width = 0.20\textwidth]{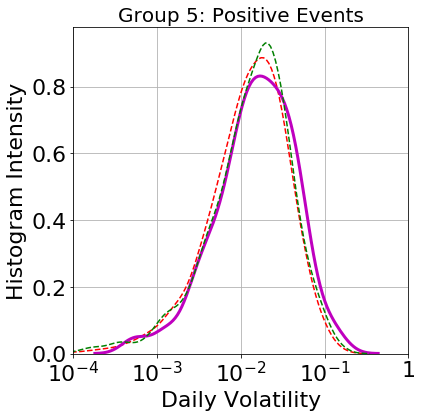}}
  
  \includegraphics[width = 0.4\textwidth]{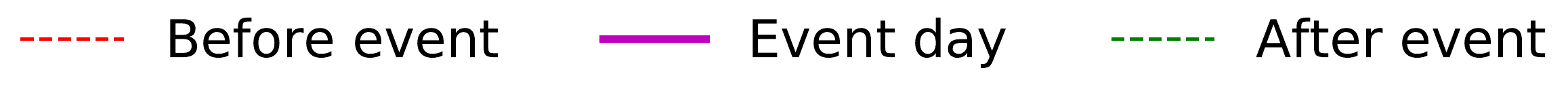}
  \caption{Probability density function of \textbf{AR} and \textbf{daily realised volatility} of Group 2 companies, computed over all group-aggregated sentiment events over the time period: (a) \textbf{AR} around \textbf{negative events}, (b) \textbf{AR} around \textbf{positive events}, (c) \textbf{Volatility} around \textbf{negative events}, (d) \textbf{Volatility} around \textbf{positive events}.}
  \label{fig:group2dist}
\end{figure}

\section*{Discussion}

In this work, we have taken a first step in integrating NLP-based financial news sentiment analysis and network analysis of financial markets. In particular, we propose a novel pipeline that dynamically analyses the evolution of stocks of major companies using a network constructed on the basis of financial news and sentiments. We extend this framework to analyse the inter-connectedness of sentiment by considering the correlation of sentiment of individual companies to groups of related companies. We also demonstrate the practical utility of this network representation, as we find a non-negligible market movement associated with sentiment beyond the individual company level. 

We believe that the present study points to an exciting direction of future research to improve the understanding of financial markets, the ultimate goal of this work. Up to this point, analysis of media sentiment and market movements in the existing literature has been largely focused on individual companies or financial assets. However, our work provides evidence suggesting that 
taking into account the dynamic interplay between these companies or assets
and a shift from single company to at least sector/group level would be beneficial. We believe that this better understanding and quantification of sentiment correlation (which has been largely neglected) will be crucial, and could lead to better financial decision-making at various levels, from market participants to market regulators. 
Specifically, for market participants, a deeper understanding of news sentiment and market movement beyond the level of individual companies may help them make more informed investment decisions. For market regulators, insights about the collective dynamics of news co-occurrence, media sentiment, and associated market movement may provide them with a more holistic picture of the financial market, and allows for the monitoring of systemic risk in the market as well as designs of effective intervention mechanisms.
Finally, apart from the empirical findings based on retrospective analysis, the pipeline proposed in the paper is highly valuable in its own right in that it provides a real-time network analysis tool on sentiment and market movements.

Nonetheless, it is important to also bear in mind the limitations of our present study.
First, we only analyse news articles from one particular outlet (i.e. Reuters), which might subject us to hidden bias and scarcity of data. To partially ameliorate the second effect, we limit our focus to the largest companies that receive an adequate amount of coverage from Reuters news. 
Future work might consider extending the scope by including a broader range of companies and markets. 
Second, we are also limited by the present state-of-the-art of NLP analysis on effective entity recognition and sentiment prediction, which remain open research questions in the NLP community. In this aspect, the most obvious next step would be to employ  more advanced techniques once they become available, and to use a more domain-specific corpus related to finance jargon, for example, for model training. 
Third, in the present work, we focus on the \textit{correlation} of sentiments and associated market movements based on the historical data. In particular, no analysis of \textit{causality} has been attempted; nevertheless, we note that, in many cases, the associated market movements took place \textit{after} the sentiment movements (for example, see Figure \ref{fig:individualmarketimpact} -- while the largest movement in log-return is often seen on event days, the same trend can often be extended to a longer period of time after the event day). Therefore, an immediate extension work would be to investigate any causal link between sentiment and market movement, a study we believe could be potentially extremely useful for financial decision-making.
Fourth, the current analysis of sentiment effect ignores some of the possible interfering effects. For example, we ignore overlaps of multiple sentiment events from different companies which might be highly related. 
Future work could seek to rectify this by devising methods to 
control for such possible interfering effects.
Finally, we acknowledge that, for the ease of analysis and data acquisition, we have only looked at historical data, and we hope to extend the study to more recent period of time, especially in view of the recent market turbulence and the possible changing market landscape in recent years due to increased automation \cite{amaro_2018}. 

\section*{Methods}

We focus on the investigation of network relations between companies through NLP-based analysis of financial news text. 
We first collect the raw text of financial news and then use a named entity recognition system to automatically identify the organisations in the financial news. The recognized organizations were cleaned and 87 target companies were kept in a post-processing step (as detailed below). A state-of-the-art sentiment prediction model is used to predict the sentiment score of those identified companies when they appear in the news.  
This allows us to build time series of company sentiment and study its dynamics and market movement using the news co-occurrence network of the companies.

\subsection*{Data sets}
We use the financial news on Reuters collected from October 2006 to November 2013, which are publicly available \cite{ding2014using}. Table \ref{tab:statistics} shows some statistics of the data set. We select the data from January 1st, 2007 to September 30th, 2013 to include 27 full quarters. For market data, we collect daily closing price and volatility data of 87 target companies (methods to identify these companies are outlined in the next section) using the Bloomberg Terminal \cite{bloomberglp}.

\begin{table}[h]
\begin{center}
\caption{Statistics of datasets.}
\begin{tabular}{|l|l|l|l|l|}
\hline
\textbf{Date} &\textbf{\#Days}&\textbf{\#News} & \textbf{\#Sentences} & \textbf{\#Words} \\ 
\hline
2007.01.01-2007.12.31 &365& 11,180 &218,969   & 5,645,015  \\ 
2008.01.01-2008.12.31 &366& 15,422 &331,543   & 8,575,485  \\ 
2009.01.01-2009.12.31 &365& 12,714 &295,406   & 7,654,841  \\ 
2010.01.01-2010.12.31 &365& 11,660 &302,802   & 7,813,895  \\ 
2011.01.01-2011.12.31 &365& 14,037 &368,330   & 9,664,028  \\ 
2012.01.01-2012.12.31 &366& 16,813 &419,741   & 10,993,426  \\
2013.01.01-2013.09.30 &272& 11,030 &259,570   & 6,900,669   \\ 
\hline
Total &2,464&92,856&2,196,361&57,247,359 \\
\hline
\end{tabular}
\label{tab:statistics}
\end{center}
\end{table}

\subsection*{Named Entity Recognition}
We use a state-of-the-art deep learning based named entity recognition (NER) system NCRF++ \cite{yang2018ncrf} to automatically extract entities from financial news text. We use a character-level convolutional neural network (CNN) \cite{lecun1989backpropagation} with a word-level long short-term memory (LSTM) \cite{hochreiter1997long} network to extract the text features and use the conditional random field (CRF) \cite{lafferty2001conditional} as the decoding layer. The NER system is trained on the CoNLL 2003 dataset \cite{tjong2003introduction}. Only the identified entities with type ``organisation'' are kept for the following steps.

As one organisation may have various name formats (e.g., ``Apple Inc.'',``AAPL'',``Apple'' all refer to the same organisation), we normalise the identified organisation names using the following rules: 1) we manually set a number of criteria to aggregate entities with short and full names. For example, ``XXX LLC'' $\rightarrow$ ``XXX'' and ``XXX Group'' $\rightarrow$ ``XXX''; 2) we automatically disambiguate entities through brackets. In more detail, we utilise the bracket information within the news text and extract the mapping pairs. For example,  ``expanded partnership with International Business Machines Corp (IBM)'' implies that ``IBM'' is the abbreviation of text ``International Business Machines Corp''. 

We keep the organisations that are consistently mentioned more than 4 times in the news in each of the 27 quarters, which resulted in 145 frequent organisations. Among those frequent organisations, we select 87 companies with a relevant price ticker in the Bloomberg terminal. We categorise the 87 companies into 9 sectors based on the Bloomberg sector list 
in 2018. Detailed company list and sector information are shown in Table \ref{tab:long}.

\subsection*{News Co-occurrence Network}
In this paper, we conduct a series of investigations on the News Co-occurrence Network. Firstly, we perform \textit{Community Detection} on the network to obtain clusters of related companies:
Given the news co-occurrence network constructed from the data of the first year, we apply the \textit{Louvain modularity} method \cite{blondel2008fast} to detect the communities of the related companies which we term the ``groups'' in this paper. We present a high-level overview of the Louvain modularity method as follows. The Louvain method uses \textit{modularity} as the objective function $Q$ it aims to maximise: 
\begin{equation}
    Q = \frac{1}{2m}\sum_{i, j}\Big( e_{i, j} - \frac{k_i k_j}{2m}\delta(C_i, C_j) \Big),
\end{equation}
where the summation is over all edges in the network, $e_{i, j}$ is the weight of the edge connecting nodes $i$ and $j$ (in our case, the cosine similarity measure between companies in terms of news appearance), $k_i$ and $k_j$ are the sum of all weights of the edges attached to nodes $i$ and $j$, respectively, $c_i$ and $c_j$ are the communities $i$ and $j$ belong to (in our case, the company groups), respectively, and $\delta$ is the Kronecker delta function with $\delta(x, y) = 1$ if $x = y$ and $\delta(x,y)=0$ otherwise. The Louvain method initialises by assigning each node its own community and then computes the change in modularity, $\Delta Q$, by removing node $i$ from its own community and moving it into each community $i$ is connected to. Once computed for all communities, the node $i$ is assigned to the community that \textit{leads to largest $\Delta Q$}, if any increase $\Delta Q$ is possible (otherwise the community of node $i$ remains unchanged). This process is repeated sequentially for all nodes until no further $\Delta Q > 0$ is possible.

After concluding the aforementioned procedures the algorithm starts the second phase, where nodes of the same community are now represented as \textit{nodes in a new network} and the first phase can be re-applied again. This two-phased \textit{pass} continues until there is no change in the computed communities and the algorithm terminates. We use the Python implementation of the Louvain modularity method (\url{https://github.com/taynaud/python-louvain}), which further complements the aforementioned Louvain modularity method with the multiscale feature\cite{lambiotte2008laplacian} -- in this work, we set the timescale parameter associated with this feature to the default value of $1.0$. 

\begin{figure}[t]
  {\includegraphics[width = 0.4\textwidth]{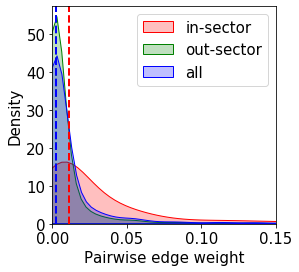}}
  \caption{Probability distributions of pairwise edge weights between companies in the co-occurrence networks of companies belonging to the same sector (in-sector), companies belonging to the different sectors (out-sector) and overall distributions (all). The dotted vertical lines denote the population medians for each distribution. Note that the \textit{out-sector} and \textit{all} medians overlap because most edges are out-sector, since it is more likely for two arbitrary companies to belong to different sectors than the same sector.}   
  \label{fig:weightdistr}
\end{figure}

Given the clusters of related companies obtained from the aforementioned procedures, we also compare and contrast these clusters with the ground-truth \textit{sectors} provided by Bloomberg, and we are particularly interested in isolating the interesting pairs of companies where the clusters reveal relations \textit{different} from the sector information, or otherwise provides more informative insights. Given that we find company pairs belonging to the same sector are also more likely to have stronger weights amongst them (and hence more likely to appear in the same cluster. See Figure \ref{fig:weightdistr}), we design filtering criteria that explicitly take account of the different weight distributions of the companies that belong to the same sectors (\textit{in-sector}) and those do not (\textit{out-sector}). We isolate the outlier companies pairs defined as those with edge weights above 75th percentile + 1.5 Interquartile Range (IQR) in both categories.

\subsection*{Sentiment Prediction}
The identification of sentiment polarity for a given entity in a context is a classical ``targeted sentiment analysis'' NLP task \cite{jiang2011target}. Given the identified entities, we use a state-of-the-art attentive neural network model to predict the sentiment of given entities \cite{liu2017attention}. The model was trained using the text corpus made publicly available from a previous study \cite{zhang2016gated}. For each sentence with given entities, our sentiment model assigns each entity a sentiment value ranging from -1 to 1, where -1 represents the most negative polarity and 1 the most positive one. 
Our sentiment model utilises the full context information of the sentence in assigning sentiment value to the target entity.

One entity may have been mentioned multiple times in the news during the period of analysis
(e.g., one day or one quarter). In this case, we use the average sentiment among all mentions and we define a news article to be \textit{sentiment-bearing} if it contains an overall non-neutral sentiment towards any target companies. This yields a sentiment time series of any desired temporal frequency for each of the target companies. When we considered the sentiment score at a group level, we simply aggregate the sentiment scores of all constituent entities of that group by averaging those scores.

\subsection*{Sources of Sentiment}

As mentioned, to support the rigour of the conclusions drawn we investigate the sources of the sentiment; however, since it is infeasible to manually classify the source for each sentence, we focus on a small yet sufficiently typical sample of sentences: we first examine the distribution of the non-neutral sentiment contained in articles, as only a fraction of all the sentences in each articles carry non-zero sentiment, and it is possible to trace total amount of sentiment contained in an article back to individual contribution of these sentences. We find that the distribution of the sentiment across articles is highly skewed, with top 9.2\% articles accounting for 50\% of the sentiment directed to any of our 87 target companies (see Figure \ref{fig:cumulativesentiment} in Supplementary Information for details). With this showing the top few articles disproportionately account for a large amount of sentiment overall, we manually inspect all sentences with non-zero sentiment scores in the top-20 articles with most sentiments, and we find that only 1 article has a considerable amount of sentiment that is derived from market commentary. Upon visual inspection, this trend also at least extends to the top-50 articles with the most sentiment.
Furthermore, on a daily basis, we also manually look into the top-5 days with the largest magnitudes of sentiment scores (often accompanied with large price movements as well) for 3 representative companies (AAPL (Apple), GS (Goldman Sachs) and GM (General Motors)). By inspection, it is evident that most of the sentiment comes from comments on the fundamentals rather than simply commenting on the market. As a singular example, on 20 Apr 2010 Goldman Sachs (GS) experienced negative shock in sentiment. We manually screened the first 50 sentences with non-zero sentiment scores on that day, and only 4 are some sort of commentary. Even for the 4 market comments, none is directly commenting on the market performance of GS \textit{on that day}. The vast majority of the sentiment on that day directed to GS derives from the reporting a negative real-world event about GS. We find that the similar patterns hold for other days and for other companies.

\subsection*{Sentiment Event Days and Group Aggregate Sentiment}
To distill the most significant information from the sentiment time series, we extract the days where the entity or the group of entities experienced extraordinary changes in the computed sentiment scores. We denote these as \textit{event days}, which are defined as days on which the sentiment score of the entity or the group of entities exceeded 2 standard deviations above or below 
the average sentiment in the preceding 180 trading days. Based on the direction of this movement, we term the event days as \textit{positive} or \textit{negative} event days, where appropriate.

As discussed, we are primarily interested in investigating how strong sentiment in one company correlates with behaviour of related companies and it is thus important to first examine the level of correlation between the sentiment of different companies itself, and this is another reason why we focus on the sentiment events. By modelling the daily sentiment as a multivariate time series for all the companies considered, a factor analysis model with 5 latent factors (similar to the model used in Vassallo et al, 2019 \cite{vassallo2019tale}) is capable of explaining 55\% of the total variance; on the sentiment \textit{event} series (i.e. the time series with only 3 possible values: -1 (negative event), 0 (no event) and 1 (positive event)), the same model explains 2.3\%. This suggests that the strong, transient sentiments are often driven by company-specific events, instead of market- or sector-coordinated movements.

To investigate the sentiment dynamics amongst companies, we construct a news co-occurrence network as described in the main text (i.e., representing individual companies as nodes of the graph with edges being the pairwise cosine distances of vectors corresponding to the companies in the news coverage matrix). To showcase the potential predictive value of our method, we construct the networks \textit{dynamically}: on each event day described above, we only consider the news in the window of 60 trading days preceding the day itself $[T - 60, T)$. 
With this dynamic network which evolves as a function of time, for a company $c$ experiencing an event day on time $t = T$, we compute the aggregate sentiment of the companies that are top-$k$ nearest neighbours on the network measured by the edge weight; in this work, unless otherwise specified we take $k = 10$, and it is worth noting that since the networks are dynamically constructed, the nearest neighbours are in general different at a different time $t$. Formally, the sentiment score $\Bar{s_c}$ for the nearest neighbours of company $c$ at time $t = T$ is given by:
\begin{equation}
    \Bar{s}(c)|_{t=T} = \frac{1}{|N'|}\sum_{i=1}^{|N'|} s(c_i)|_{t=T}, \text{ where } c_i \in \mathrm{argmax}_{N' \subseteq N(c), |N'|=\min(k,|N(c)|)}\sum_{c_j \in N'}e_{c_j, c},
\label{eqn:sentimentscore}
\end{equation}
where $s(c_i)|_{t=T}$ is the sentiment score of company $c_i$ at time $t=T$, and $e_{c,c'}$ 
denotes the pairwise edge weight between companies $c$ and $c'$ where $c'$ in this case is a member of the neighbours $N(c)$ of company $c$ in the network. To compare the group sentiment evolution around sentiment days, $\Bar{s_c}$ is then computed for each day in the range of $[T-7, T+7]$ around the event days to produce $\{\Bar{s_c}|_{t=T-7}, ..., \Bar{s_c}|_{t=T}$\text{ (event day) }$, ... \Bar{s_c}|_{t=T+7}\}$. Notice that there is one such series defined on each of the event days for each of the companies that experienced at least one event day. We first average over \textit{all event days} to obtain one series per company. For the sake of better presentation, we then further aggregate by averaging over the \textit{groups} of the companies (see Table \ref{tab:clusterfirstyear}) to condense 87 series to 7, as we expect the companies in the same group to behave more similarly. Formally, following the notation in equation (\ref{eqn:sentimentscore}), the final \emph{group aggregate sentiment score} of a group of companies $G$, $s(G)$, which is the quantity represented in Figure \ref{eqn:sentimentscore} and discussed in the preceding section, can be mathematically represented by the two level aggregation:
\begin{equation}
    s(G)|_{t=\tau} = \frac{1}{|G|}\sum_{c \in G} \Big( \frac{1}{|E(c)|}\sum_{e \in E(c)}\Bar{s}(c)|_{t = e + \tau}\Big), \text{ } \forall \tau \in [-7, 7],
\end{equation}
where $E(c)$ is the set of event days of company $c$ over the period of time considered. It is worth emphasising that this quantity reflects the group sentiment as a whole and is independent of timestamp $T$ and is only dependent on $\tau$, which is the the number of days \textit{relative} to the sentiment day.

While it is infeasible to conduct the above screening procedure for all articles considering the vast number of sentences contained, we argue that at least from a typical sample, most sentiment,  \textit{at least on event days where there are massive changes in sentiment}, is primarily derived from underlying fundamental events, instead of trivially mirroring the market performance. We argue this is also reasonable: qualitatively, while Reuters does publish some articles purely for market commentary, we do expect the majority of the daily articles especially on days with important breaking news to be dedicated to reporting the underlying events, which should be the primary drivers of both the market and the sentiment. For the latter category of articles, the market performance is usually only commented on sparingly, and it is thus expected that the sentiment is primarily driven by the underlying events.

\subsection*{Market Data Processing}
To accurately quantify the market movements associated with sentiment, we process the market data to exclude other possible confounding factors. Specifically, we compute the \textit{cumulative abnormal return} (CAR) in the 7 trading days leading up to and immediately after the sentiment event days, defined in the previous section. Consider an event day on day $T$, on an arbitrary day around that day $t$, the CAR is given by:
\begin{equation}
    CAR(t) = \sum^{t} _{i = t-T_s} \epsilon_i,
\end{equation}
\noindent where $\epsilon_t$ denotes he abnormal return (AR) on day $t$, and the summation is over $t-T_s+1$ trading days, where $T_s$ is set to a fixed value of $7$ trading days before the event day. 

The AR is the excess return over the expected return from the Capital Asset Pricing Model (CAPM). Here we follow Ranco \textit{et al.}\cite{ranco2015effects} to select the CAPM model\cite{fama2004capital} over the more popular Fama-French model\cite{fama1992cross, fama2015five}: this is because we both favour the simplicity of CAPM model and expect that the additional factors included in Fama-French model (i.e. the factors explaining the outperformance of small-cap over large-cap companies, and that of value stocks over growth stocks) would be largely consistent over our selection of companies, which are predominantly large-cap and usually fall to the ``growth'' bracket. The market model decomposes the return of a single stock at time $t$, $R_t = \log\Big(\frac{P(t)}{P(t-1)}\Big)$ ($P(t)$ is simply the closing price at day $t$), with three components in a linear manner: $\beta$, which captures the return that can be explained by the movement of the entire market (usually represented by the index log-return $R_m$); $\alpha$, which is the idiosyncratic return over the index; and $\epsilon$, a stochastic term to explain any residual influence that is the AR in our context:
\begin{equation}
    R_t = \alpha_t + \beta_t R_{m} + \epsilon_{t}.
\end{equation}
\noindent Here, we compute parameters $\alpha$ and $\beta$ for each stock on each trading day based on its performance as compared to the index return in the previous 180 trading days. 
We select the MSCI World Index since the list of our selected companies include large-cap companies over a global range of developed markets. The readers are referred to Table \ref{tab:long} for a detailed list of companies and their corresponding sector listing.
The null hypothesis is based on the market model that any return of a single stock can be explained by $\alpha$ and $\beta$, and $\epsilon$ is assumed to be a zero-mean stochastic quantity. Any $\epsilon$ that significantly differs from 0 suggests there exist factors affecting the stock return that are not explained by the market model; one possible such factor is, in our case, movement in the sentiment of the company to which the stock belongs, which influences this particular stock only but not the index (at least to a lesser extent). 

To model daily volatility, we use the absolute value of the daily log-return as a proxy $\sigma(t) \approx |\log\frac{P(t)}{P(t-1)}|$. While this deviates from the exact definition of volatility as the standard deviation of the log-return, this approximation has been shown empirically to correlate strongly with the actual volatility \cite{longin2001extreme} and has been commonly used in the literature \cite{saha2009clustering, masoliver2006multiple}. This metric therefore measures the magnitude of market reaction, regardless of the direction of the response.

As a final modelling step, given the market movement series $\{CAR|_{t=T-7}, ..., CAR|_{t=T+7}\}$ and $\{\sigma|_{t=T-7}, ..., \sigma|_{t=T+7}\}$ for each company around each event day similar to the group aggregate sentiment score described in the preceding section, we apply the similar final aggregation steps to obtain one series per company group. While descriptive statistics like median and mean are usually sufficient to model the market movements on an individual company level, in a group level, we also consider the distributions of AR and volatility approximated by \textit{histograms} and \textit{kernel density estimation} (KDE). Specifically, within each group defined in Table \ref{tab:clusterfirstyear}, we first compute a group level sentiment time series by aggregating the sentiment time series of all the members of the group. Next, on market data, without aggregation, we categorize the elements of the CAR and volatility series of the related companies into \textit{before event day} $t \in [T-7, T)$, \textit{on event day} $t = T$, \textit{after event day} $t \in (T, T+7]$. Within each category, over the entire period of time considered, we obtain the histograms of AR and volatility - by the market model described in the preceding section, $AR \sim \mathcal{N}(0, \sigma^2)$. With the histograms, we then apply KDE to obtain a continuous approximation $\Hat{p}(x)$ to the probability density function (PDF) of AR $p(x)$:
\begin{equation}
    \Hat{p}(x) = \frac{1}{nh}\sum_{i=1}^nK\Big(\frac{x - x_i}{h}\Big) \text{ with }  K(x) = \exp{\big(-x^2\big)}, 
\end{equation}
where $n$ is the number of data points we use to estimate the distribution, $\{x_i\}$ are the observed market data (AR or volatility) and $h$ is the bandwidth, which we estimate automatically from the Seaborn visualisation package\cite{seaborn}. 
Finally, it is worth noting that to ensure a rather accurate estimation of the distribution functions, we only apply KDE when we have more than 20 sentiment events in a particular group.

\bibliography{sample}

\clearpage
\beginsupplement
\section*{Supplementary Materials}








\begin{center}
\begin{longtable}{|l|l|l|}
\caption{Company Abbreviation Map (in alphabetical order within each sector).} \label{tab:long} \\

\hline \multicolumn{1}{|c|}{\textbf{Sector}} & \multicolumn{1}{c|}{\textbf{Abbreviation}} & \multicolumn{1}{c|}{\textbf{Company}} \\ \hline 
\endfirsthead
\multicolumn{3}{c}%
{{ \tablename\ \thetable{} -- continued from previous page}} \\
\hline \multicolumn{1}{|c|}{\textbf{Sector}} & \multicolumn{1}{c|}{\textbf{Abbreviation}} & \multicolumn{1}{c|}{\textbf{Company}} \\ \hline 
\endhead
\endfoot
\hline 
\endlastfoot
\hline
\multirow{25}*{Financials (25)}&AXP&American Express.\\
&AIG&American International Group.\\
&BAC&Bank of America.\\
&BMO&Bank of Montreal.\\
&BK&Bank of New York Mellon.\\
&BCS&Barclays.\\
&BRKA&Berkshire Hathaway.\\
&BX&Blackstone.\\
&BNPP&BNP.\\
&CS&Credit Suisse.\\
&DBKGN&Deutsche Bank.\\
&FMCC&Federal Home Loan Mortgage Corp.\\
&GS&Goldman Sachs.\\
&HSBA&HSBC.\\
&ING&ING Groep.\\
&JPM&JP Morgan.\\
&KKR&KKR.\\
&LEHMQ&Lehman Brothers.\\
&MA&MasterCard.\\
&MS&Morgan Stanley.\\
&RY&Royal Bank of Canada.\\
&RBS&Royal Bank of Scotland.\\
&SOCGEN&Societe Generale.\\
&UBSN&UBS.\\
&WFC&Wells Fargo.\\
\hline
\multirow{16}*{Technology (16)}&AAPL&Apple.\\
&BBRY&BlackBerry.\\
&CEOS&CeCors.\\
&CSCO&Cisco.\\
&ERICB&Ericsson.\\
&GARTNER&Gartner.\\
&GOOG&Google.\\
&HPQ&Hewlett Packard.\\
&IBM&IBM.\\
&INTC&Intel.\\
&MSFT&Microsoft.\\
&MCO&Moody 's Inc.\\
&NOK&Nokia.\\
&SSNLF&Samsung.\\
&SNE&Sony.\\
&YHOO&Yahoo.\\

\hline
\multirow{4}*{Materials (4)}&AA&Alcoa.\\
&BLT&BHP.\\
&ISM&Inspiration Mining.\\
&RIO&Rio Tinto.\\

\hline

\multirow{5}*{Energy (5)}&BP&BP.\\
&CVX&Chevron.\\
&RDSA&Royal Dutch Shell.\\
&EC&Ecopetrol.\\
&XOM&Exxon Mobil.\\

\hline
\multirow{2}*{Communications (2)}&VZ&Verizon Communications.\\
&VOD&Vodafone.\\

\hline
\multirow{2}*{Healthcare (2)}&JNJ&Johnson \& Johnson.\\
&PFE&Pfizer.\\
\hline

\hline
\multirow{6}*{Consumer Staples (6)}
&KO&Coca-Cola.\\
&COST&Costco Wholesale.\\
&PEP&Pepsi.\\
&PG&Procter \& Gamble.\\
&TGT&Target Corp.\\
&WMT&Wal-Mart.\\

\hline
\multirow{20}*{Consumer Discretionary (20)}&AMZN&Amazon.\\
&BBY&Best Buy.\\
&BMWG&BMW.\\
&DAIGN&Daimler AG.\\
&DAL&Delta Airlines.\\
&EBAY&Ebay.\\
&FIA&Fiat Chrysler.\\
&GPS&Gap Inc.\\
&GM&General Motors.\\
&HD&Home Depot.\\
&HMC&Honda Motor.\\
&JCP&JC Penney.\\
&MCD&McDonald.\\
&NISSAN&Nissan.\\
&RENA&Renault.\\
&SBUX&Starbucks.\\
&TM&Toyota Motor.\\
&TVS&TVS Motor.\\
&UAL&United Continental Holdings.\\
&VOWG&Volkswagen AG.\\

\hline
\multirow{7}*{Industrials (7)}&EAD&Airbus SE.\\
&BA&Boeing.\\
&CAT&Caterpillar.\\
&FDX&FedEx.\\
&GE&General Electric.\\
&UPS&United Parcel Service.\\
&UTX&United Technologies.\\

\end{longtable}
\end{center}

\clearpage

\begin{table}[t]
\caption{Groups of companies based on the community detection on the news co-occurrence network of the first year ((in alphabetical order within each sector).}
\begin{center}
\label{tab:clusterfirstyear} 

\begin{tabular}{|p{2cm}|p{8cm}|p{1.5cm}|}
\hline 
\textbf{Group Number} & \textbf{Members} & \textbf{\#Members} \\ 
\hline
1& BAC, BCS, BK, BNPP, BX, CS, DBKGN, FMCC, GS, HSBA,  ING, JPM, KKR, LEHMQ, MCO, MS, RBS, RY, SOCGEN, UBSN, UTX, WFC & 22\\
2 & BMWG, DAIGN, FIA, GM, HMC, NISSAN, RENA, TM, VOWG & 9\\
3 & AAPL, AMZN, BBY, CSCO, EBAY, EC, ERICB, GARTNER, GE, GOOG, HPQ, IBM, INTC, MA, MSFT, NOK, SSNLF, SNE, TVS, YHOO & 20\\
4 & COST, GPS, HD, JCP, TGT, WMT & 6\\
5 & AIG, BA, DAL, EAD, FDX, UAL, UPS & 7\\
6 & AXP, BBRY, BP, BRKA, CAT, CEOS, CVX, ISM, JNJ, KO, MCD, PEP, PFE, PG, RDSA, SBUX, XOM & 17\\
7 & AA, BLT, BMO, RIO, VOD, VZ & 6\\
\hline
\end{tabular}
\end{center}
\end{table}

\clearpage

\begin{table}[t]
\caption{Company pairs \textbf{not} belonging to the same ground-truth sector with edge weights above 75th percentile + 1.5 IQR in the out-sector weight distribution (Figure \ref{fig:weightdistr}). Top 50 shown in descending magnitudes of edge weight.}
\begin{center}
\label{tab:outsectoroutliers} 

\begin{tabular}{|l|l|l|l|l|}
\hline 
\textbf{Index} & \textbf{Company 1} & \textbf{Sector 1} & \textbf{Company 2} & \textbf{Sector 2}\\ 
\hline
1 & AAPL & Technology & AMZN & Consumer Discretionary \\
2 & GPS & Consumer Discretionary & TGT & Consumer Staples \\
3 & AA & Materials & DAL & Consumer Discretionary \\
4 & DAIGN & Consumer Discretionary & EAD & Industrials \\
5 & AA & Materials & UAL & Consumer Discretionary \\
6 & AMZN & Consumer Discretionary & GOOG & Technology \\
7 & AMZN & Consumer Discretionary & WMT & Consumer Staples \\
8 & JCP & Consumer Discretionary & TGT & Consumer Staples \\
9 & BRKA & Financials & KO & Consumer Staples \\
10 & BBY & Consumer Discretionary & WMT & Consumer Staples \\
11 & COST & Consumer Staples & GPS & Consumer Discretionary \\
12 & AMZN & Consumer Discretionary & TGT & Consumer Staples \\
13 & BRKA & Financials & PG & Consumer Staples \\
14 & BA & Industrials & UAL & Consumer Discretionary \\
15 & GPS & Consumer Discretionary & WMT & Consumer Staples \\
16 & GE & Industrials & GM & Consumer Discretionary \\
17 & AA & Materials & BA & Industrials \\
18 & BA & Industrials & RY & Financials \\
19 & EC & Energy & ING & Financials \\
20 & EC & Energy & MSFT & Technology \\
21 & AAPL & Technology & VZ & Communications \\
22 & BBY & Consumer Discretionary & TGT & Consumer Staples \\
23 & HD & Consumer Discretionary & TGT & Consumer Staples \\
24 & GOOG & Technology & VZ & Communications \\
25 & EBAY & Consumer Discretionary & MA & Financials \\
26 & BRKA & Financials & MCO & Technology \\
27 & MCO & Technology & SOCGEN & Financials \\
28 & CSCO & Technology & RY & Financials \\
29 & BNPP & Financials & MCO & Technology \\
30 & KO & Consumer Staples & MCD & Consumer Discretionary \\
31 & JCP & Consumer Discretionary & WMT & Consumer Staples \\
32 & BA & Industrials & DAL & Consumer Discretionary \\
33 & AMZN & Consumer Discretionary & MSFT & Technology \\
34 & HD & Consumer Discretionary & WMT & Consumer Staples \\
35 & EBAY & Consumer Discretionary & GOOG & Technology \\
36 & AA & Materials & BMO & Financials \\
37 & MCD & Consumer Discretionary & RY & Financials \\
38 & EC & Energy & INTC & Technology \\
39 & CEOS & Technology & KO & Consumer Staples \\
40 & COST & Consumer Staples & JCP & Consumer Discretionary \\
41 & MCD & Consumer Discretionary & PG & Consumer Staples \\
42 & AMZN & Consumer Discretionary & SSNLF & Technology \\
43 & ISM & Materials & RY & Financials \\
44 & AMZN & Consumer Discretionary & GARTNER & Technology \\
45 & BRKA & Financials & GE & Industrials \\
46 & FIA & Consumer Discretionary & GE & Industrials \\
47 & CAT & Industrials & MCD & Consumer Discretionary \\
48 & BBRY & Technology & RY & Financials \\
49 & CAT & Industrials & KO & Consumer Staples \\
50 & BX & Financials & YHOO & Technology \\
\hline
\end{tabular}
\end{center}
\end{table}

\clearpage

\begin{table}[t]
\caption{Company pairs belonging to the same ground-truth sector with edge weights above 75th percentile + 1.5 IQR in the in-sector weight distribution (Figure \ref{fig:weightdistr}). Top 50 shown in descending magnitudes of edge weight.}
\begin{center}
\label{tab:insectoroutliers} 

\begin{tabular}{|l|l|l|l|l|}
\hline 
\textbf{Index} & \textbf{Company 1} & \textbf{Sector 1} & \textbf{Company 2} & \textbf{Sector 2}\\ 
\hline
1 & VOD & Communications & VZ & Communications \\
2 & BLT & Materials & RIO & Materials \\
3 & MSFT & Technology & YHOO & Technology \\
4 & FDX & Industrials & UPS & Industrials \\
5 & NISSAN & Consumer Discretionary & RENA & Consumer Discretionary \\
6 & GS & Financials & MS & Financials \\
7 & BMWG & Consumer Discretionary & DAIGN & Consumer Discretionary \\
8 & KO & Consumer Staples & PEP & Consumer Staples \\
9 & TGT & Consumer Staples & WMT & Consumer Staples \\
10 & DAL & Consumer Discretionary & UAL & Consumer Discretionary \\
11 & FIA & Consumer Discretionary & GM & Consumer Discretionary \\
12 & GOOG & Technology & MSFT & Technology \\
13 & BNPP & Financials & SOCGEN & Financials \\
14 & GOOG & Technology & YHOO & Technology \\
15 & BA & Industrials & EAD & Industrials \\
16 & BX & Financials & KKR & Financials \\
17 & BCS & Financials & RBS & Financials \\
18 & BK & Financials & BX & Financials \\
19 & SSNLF & Technology & SNE & Technology \\
20 & AMZN & Consumer Discretionary & BBY & Consumer Discretionary \\
21 & HMC & Consumer Discretionary & TM & Consumer Discretionary \\
22 & AAPL & Technology & SSNLF & Technology \\
23 & DAIGN & Consumer Discretionary & FIA & Consumer Discretionary \\
24 & MCD & Consumer Discretionary & SBUX & Consumer Discretionary \\
25 & DAIGN & Consumer Discretionary & RENA & Consumer Discretionary \\
26 & CS & Financials & UBSN & Financials \\
27 & AAPL & Technology & GOOG & Technology \\
28 & COST & Consumer Staples & WMT & Consumer Staples \\
29 & HMC & Consumer Discretionary & NISSAN & Consumer Discretionary \\
30 & AMZN & Consumer Discretionary & EBAY & Consumer Discretionary \\
31 & GARTNER & Technology & NOK & Technology \\
32 & GS & Financials & JPM & Financials \\
33 & AXP & Financials & MA & Financials \\
34 & HPQ & Technology & IBM & Technology \\
35 & RDSA & Energy & XOM & Energy \\
36 & AAPL & Technology & GARTNER & Technology \\
37 & CVX & Energy & XOM & Energy \\
38 & NISSAN & Consumer Discretionary & TM & Consumer Discretionary \\
39 & BMWG & Consumer Discretionary & VOWG\_P & Consumer Discretionary \\
40 & DAIGN & Consumer Discretionary & VOWG\_P & Consumer Discretionary \\
41 & BAC & Financials & WFC & Financials \\
42 & GARTNER & Technology & MSFT & Technology \\
43 & NOK & Technology & SSNLF & Technology \\
44 & BCS & Financials & LEHMQ & Financials \\
45 & MSFT & Technology & NOK & Technology \\
46 & BAC & Financials & GS & Financials \\
47 & COST & Consumer Staples & TGT & Consumer Staples \\
48 & CS & Financials & GS & Financials \\
49 & AAPL & Technology & NOK & Technology \\
50 & BAC & Financials & JPM & Financials \\
\hline
\end{tabular}
\end{center}
\end{table}

\clearpage
\floatsetup[figure]{subcapbesideposition=top}

\clearpage
\floatsetup[figure]{subcapbesideposition=top}

\clearpage
\floatsetup[figure]{subcapbesideposition=top}
\begin{figure}[t]
  {\includegraphics[width = 0.4\textwidth]{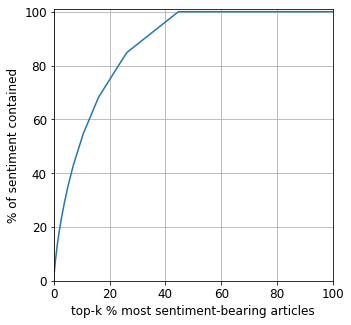}}
  \caption{Fraction of total non-neutral sentiment (i.e. either positive or negative sentiment) towards the 87 target companies versus the fraction of non-neutral sentiment-bearing articles. It is clear that the a small number of articles account for a disproportionately large amount of sentiment directed to the target companies, with top 9.2\% articles accounting for 50\% sentiments and top 32.6\% articles accounting for 90\% sentiments. Slightly more than half of the articles do not contain non-neutral sentiment to the target companies. }  
  \label{fig:cumulativesentiment}
\end{figure}

\clearpage
\floatsetup[figure]{subcapbesideposition=top}
\begin{figure}[t]
  {\includegraphics[width = 0.9
  \textwidth]{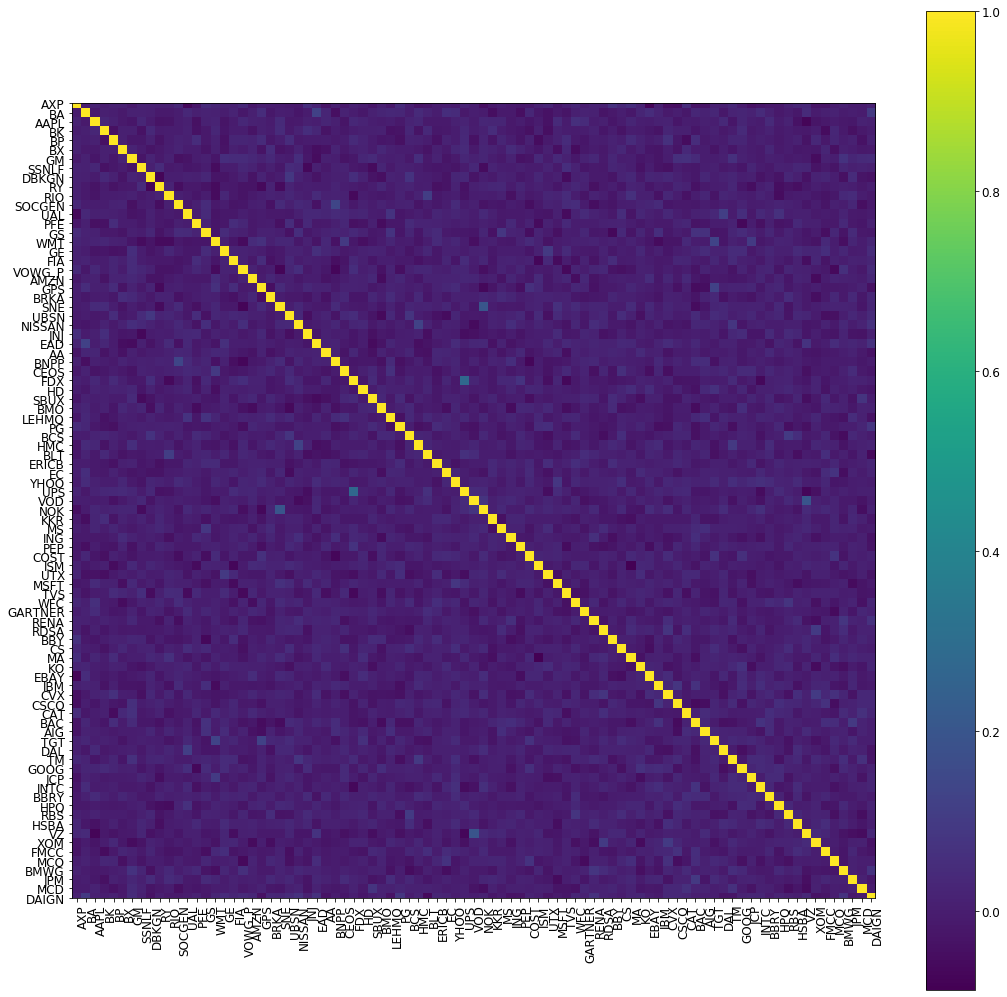}}
  \caption{Pair-wise correlation of the sentiment event time series of all the 87 target companies }
  \label{fig:sentimenteventcorr}
\end{figure}

\clearpage
\floatsetup[figure]{subcapbesideposition=top}
\begin{figure}[t]
  {\includegraphics[width = 0.8\textwidth]{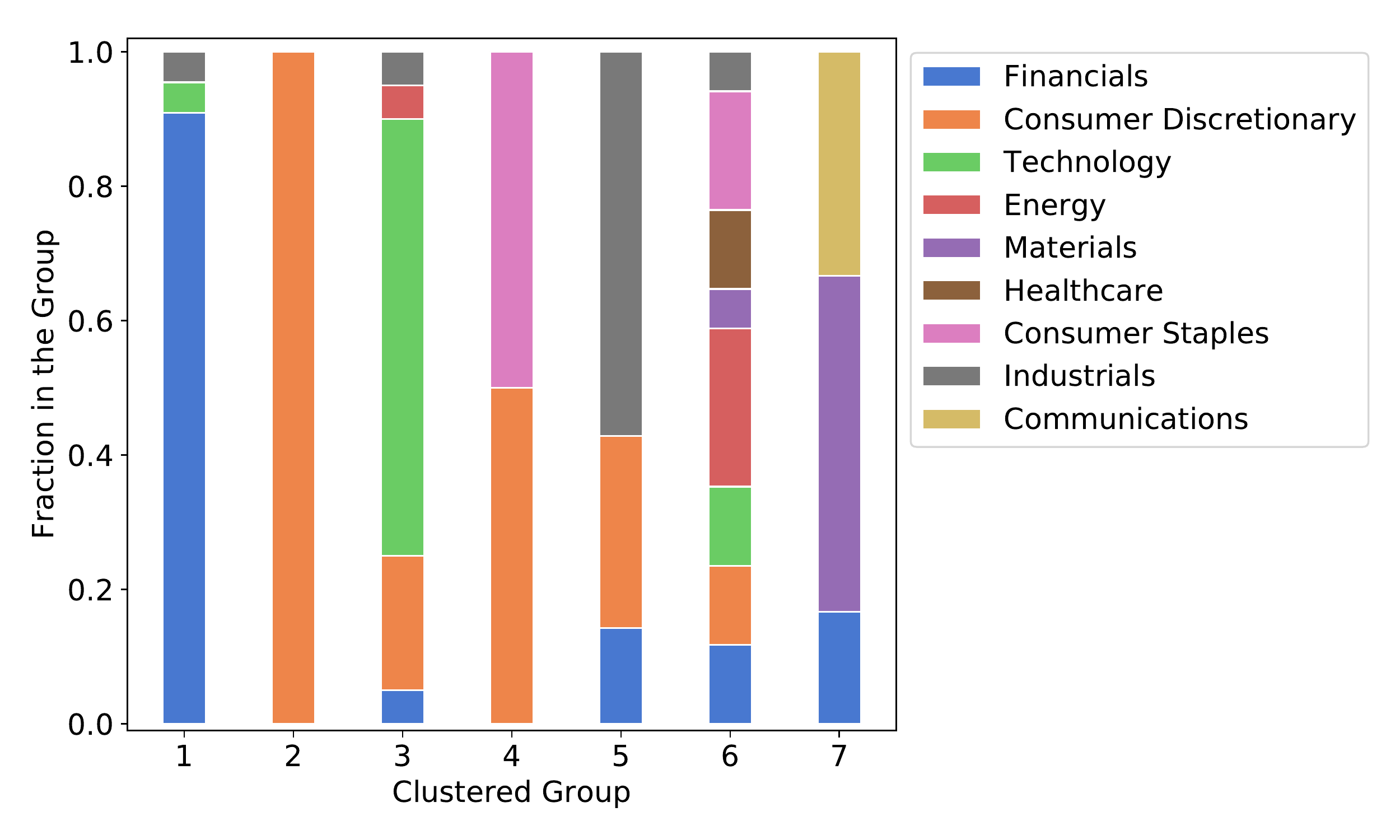}}
  \caption{Distribution of sectors in 7 clustered groups in Figure \ref{fig:general_figure}(c).}   
  \label{fig:sector_cluster}
\end{figure}

\clearpage
\floatsetup[figure]{subcapbesideposition=top}
\begin{figure}[t] 
  \subfloat[]{\includegraphics[width =0.3\textwidth]{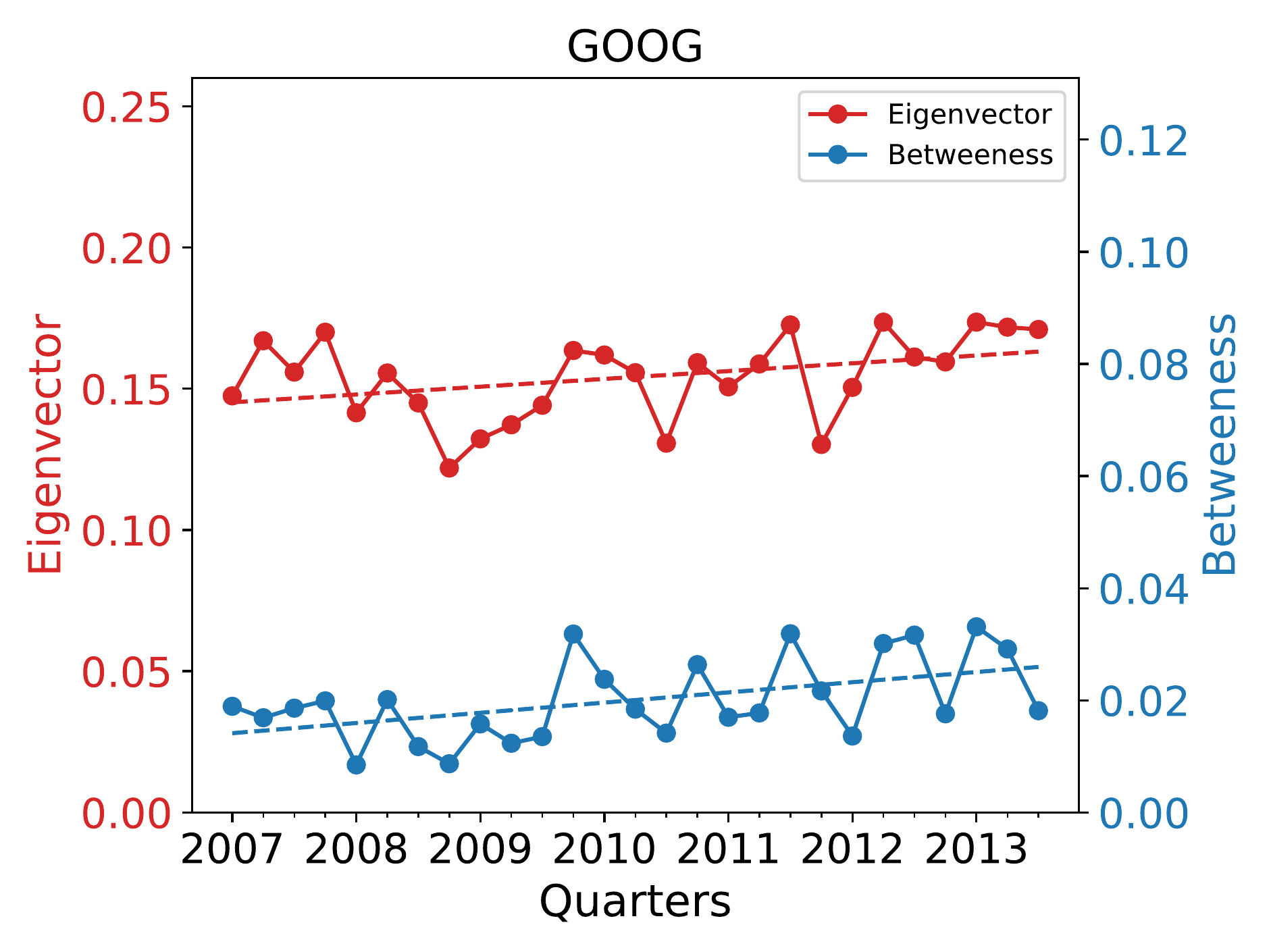}}
  \subfloat[]{\includegraphics[width =0.3\textwidth]{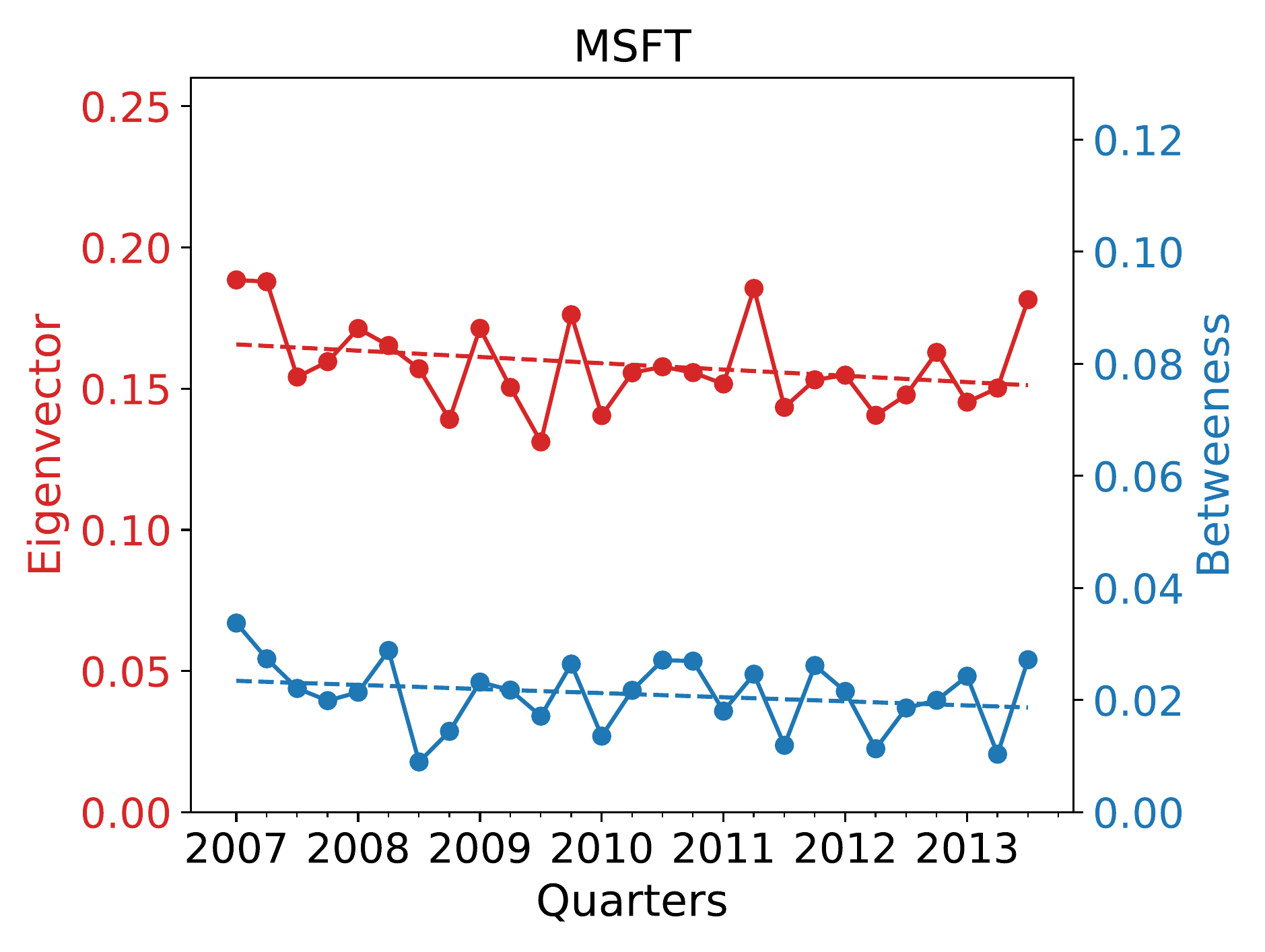}}
  \subfloat[]{\includegraphics[width =0.3\textwidth]{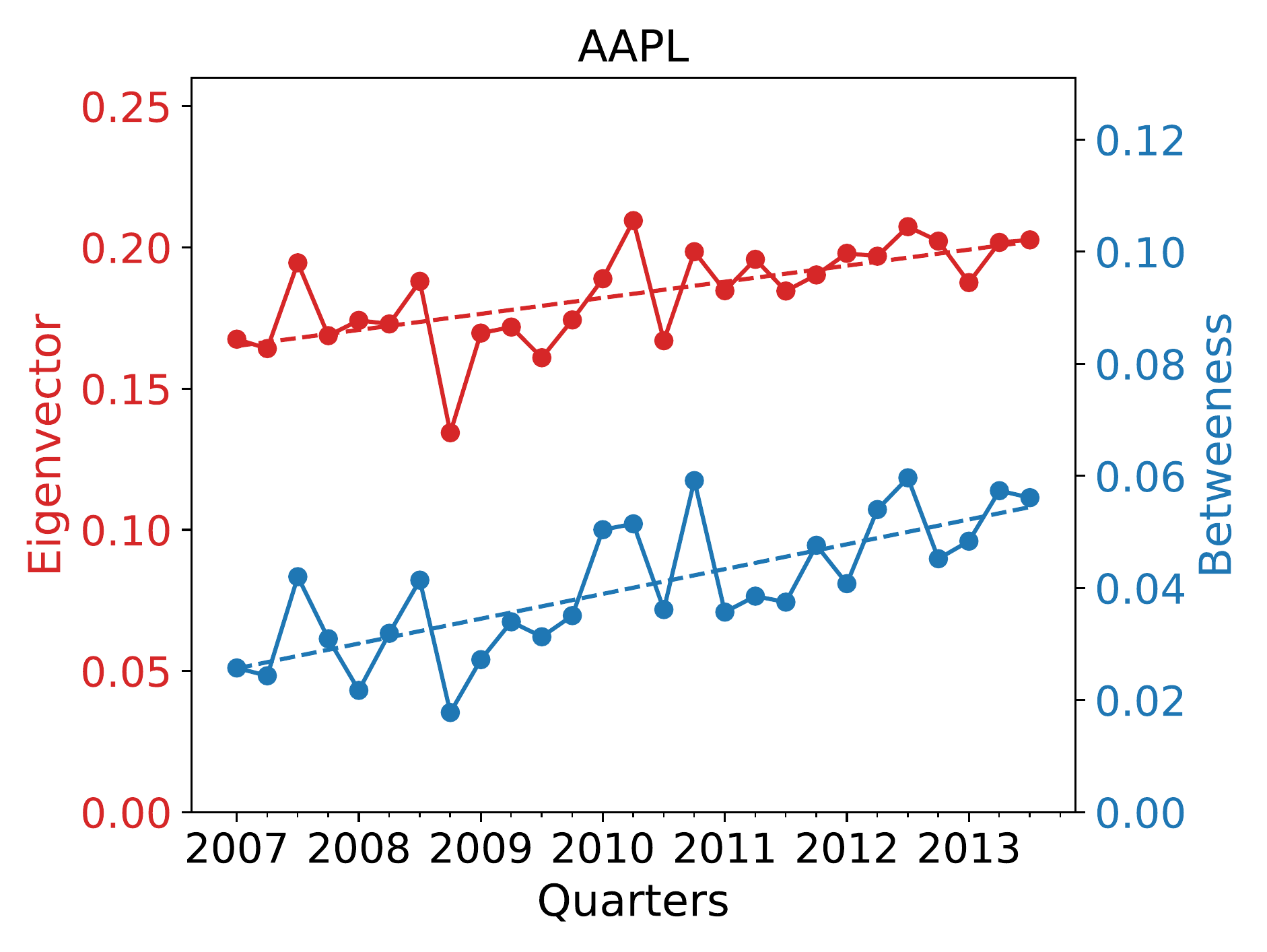}}\\
  \subfloat[]{\includegraphics[width =0.30\textwidth]{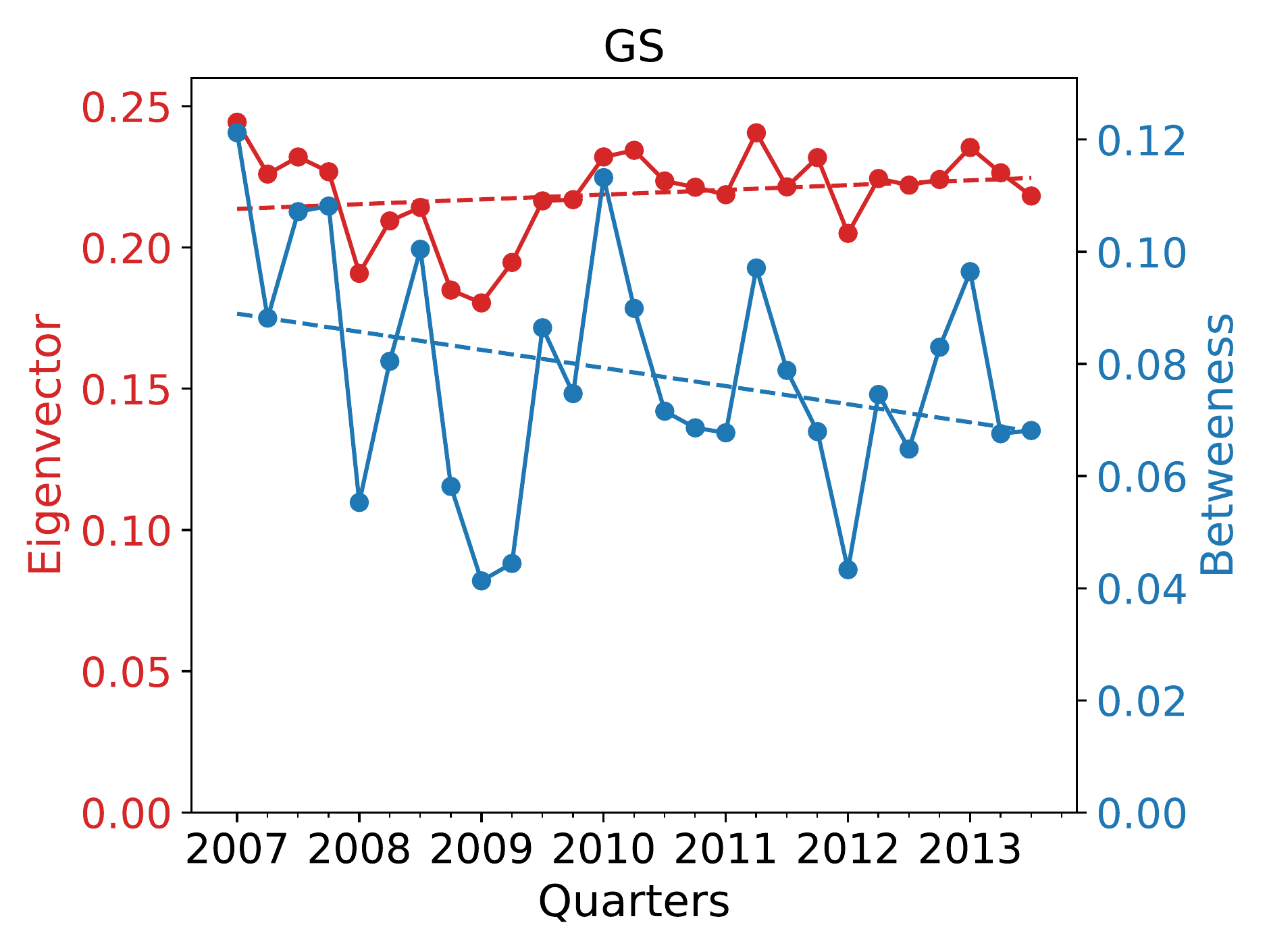}}
  \subfloat[]{\includegraphics[width =0.30\textwidth]{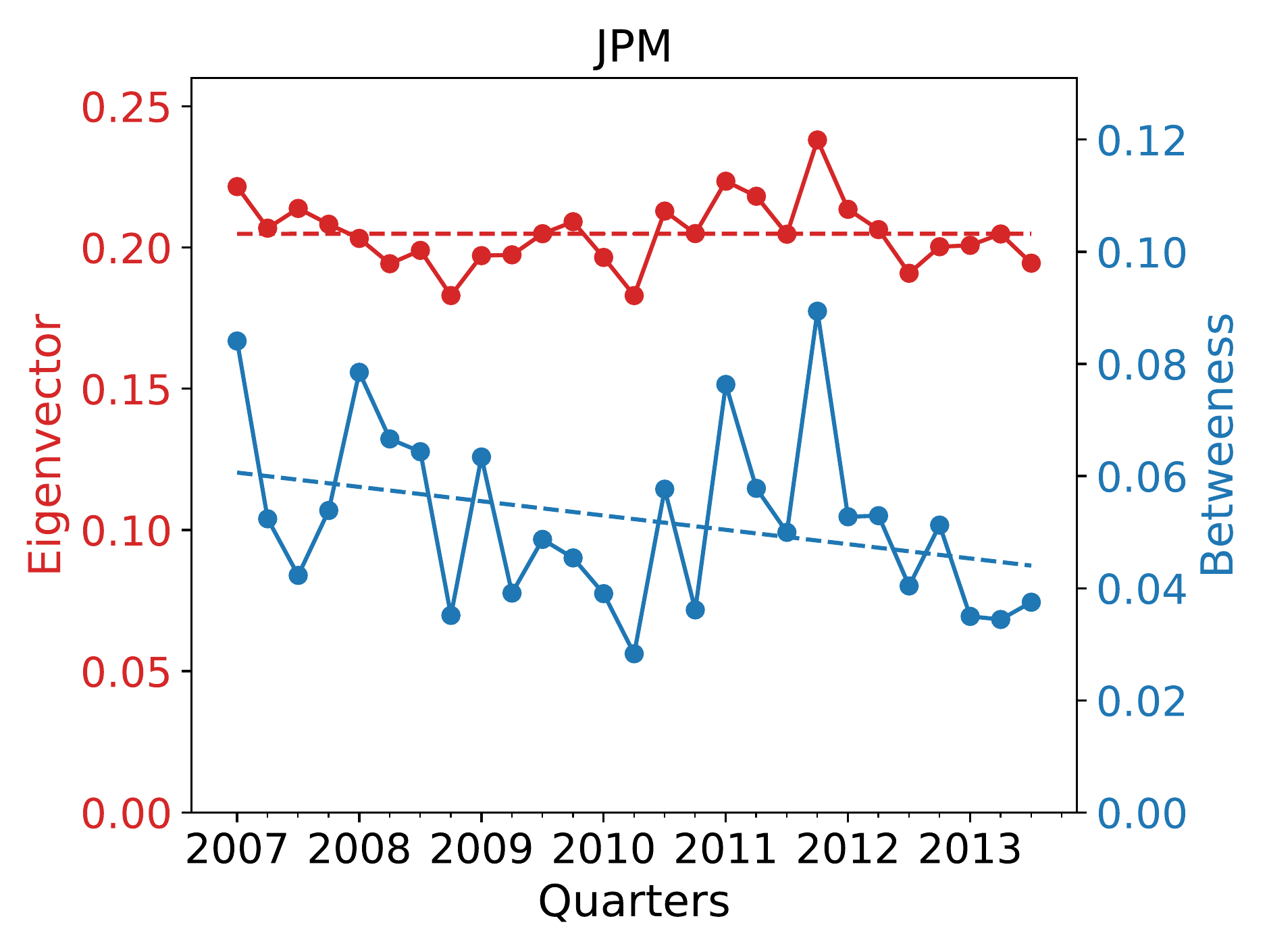}}
  \subfloat[]{\includegraphics[width =0.30\textwidth]{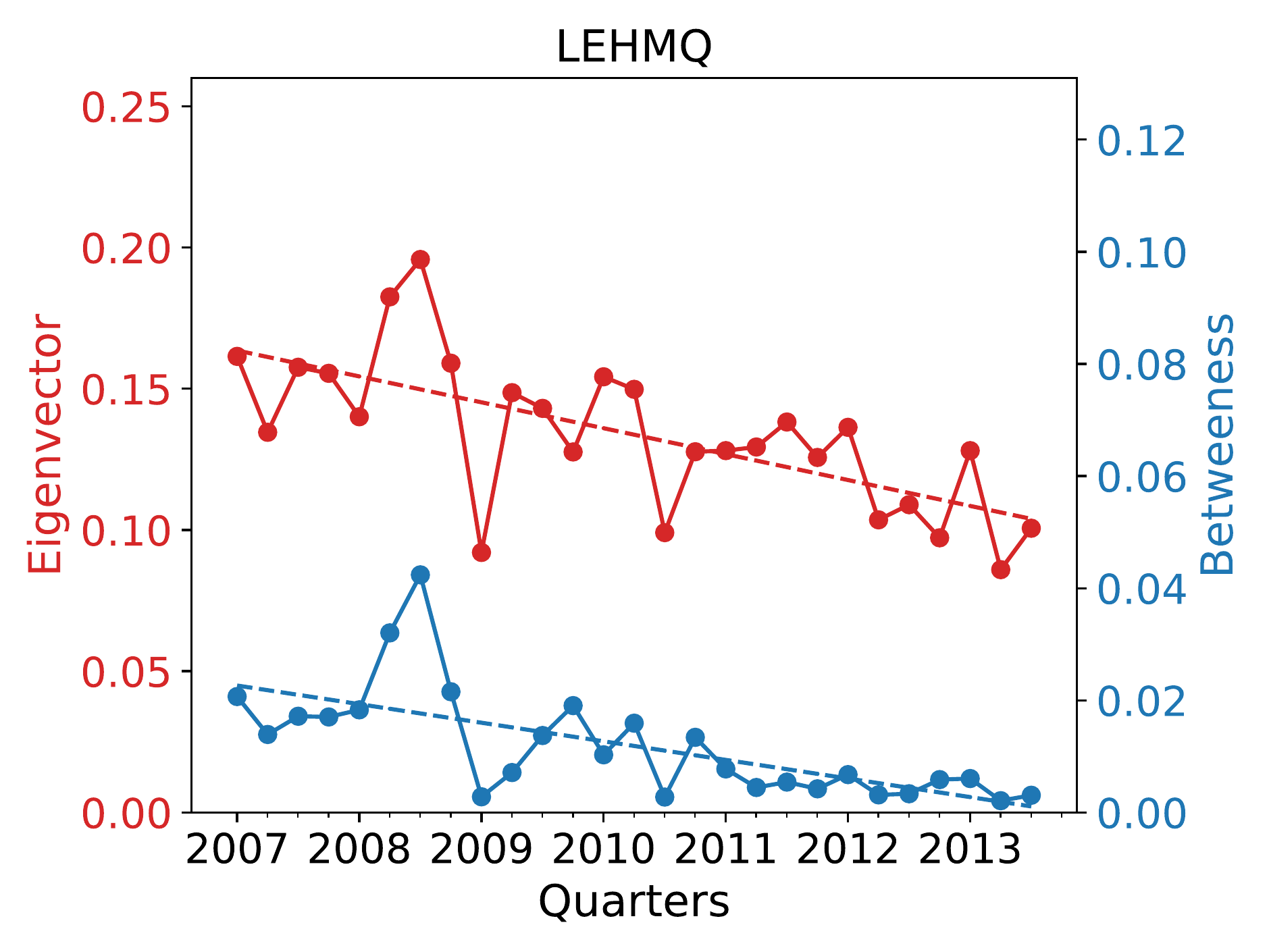}}

  \caption{Evolution of the eigenvector (red) and betweenness (blue) centrality measures of several representative companies from the ``Technology'' and ``Financial Services'' sectors: (a) Google; (b) Microsoft; (c) Apple; (d) Goldman Sachs; (e) JP Morgan; (f) Lehman Brothers.}
  \label{fig:centrality_single}
\end{figure}

\clearpage
\floatsetup[figure]{subcapbesideposition=top}
\begin{figure}[t] 
  \subfloat[]{\includegraphics[width =0.3\textwidth]{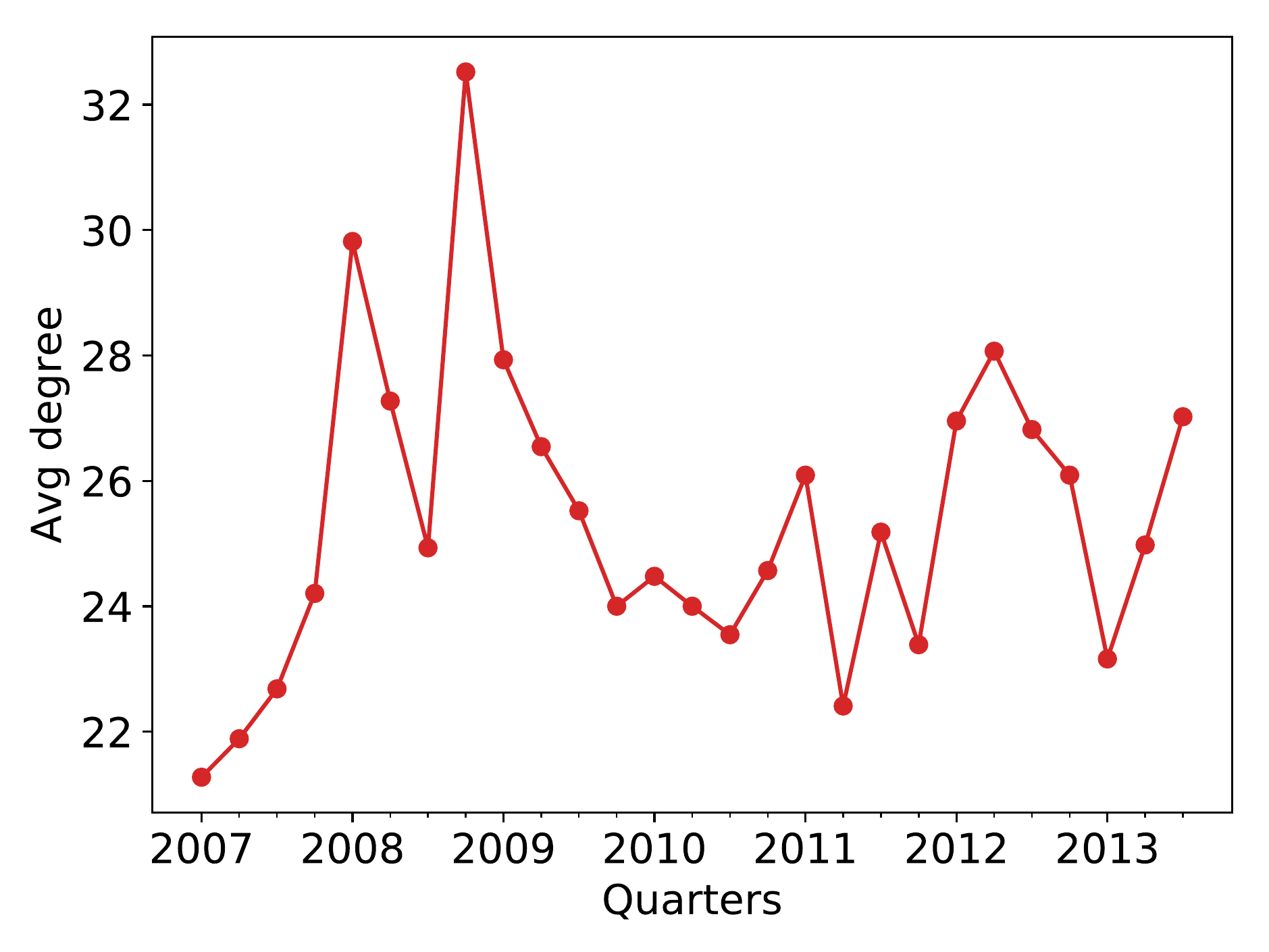}}
  \subfloat[]{\includegraphics[width =0.3\textwidth]{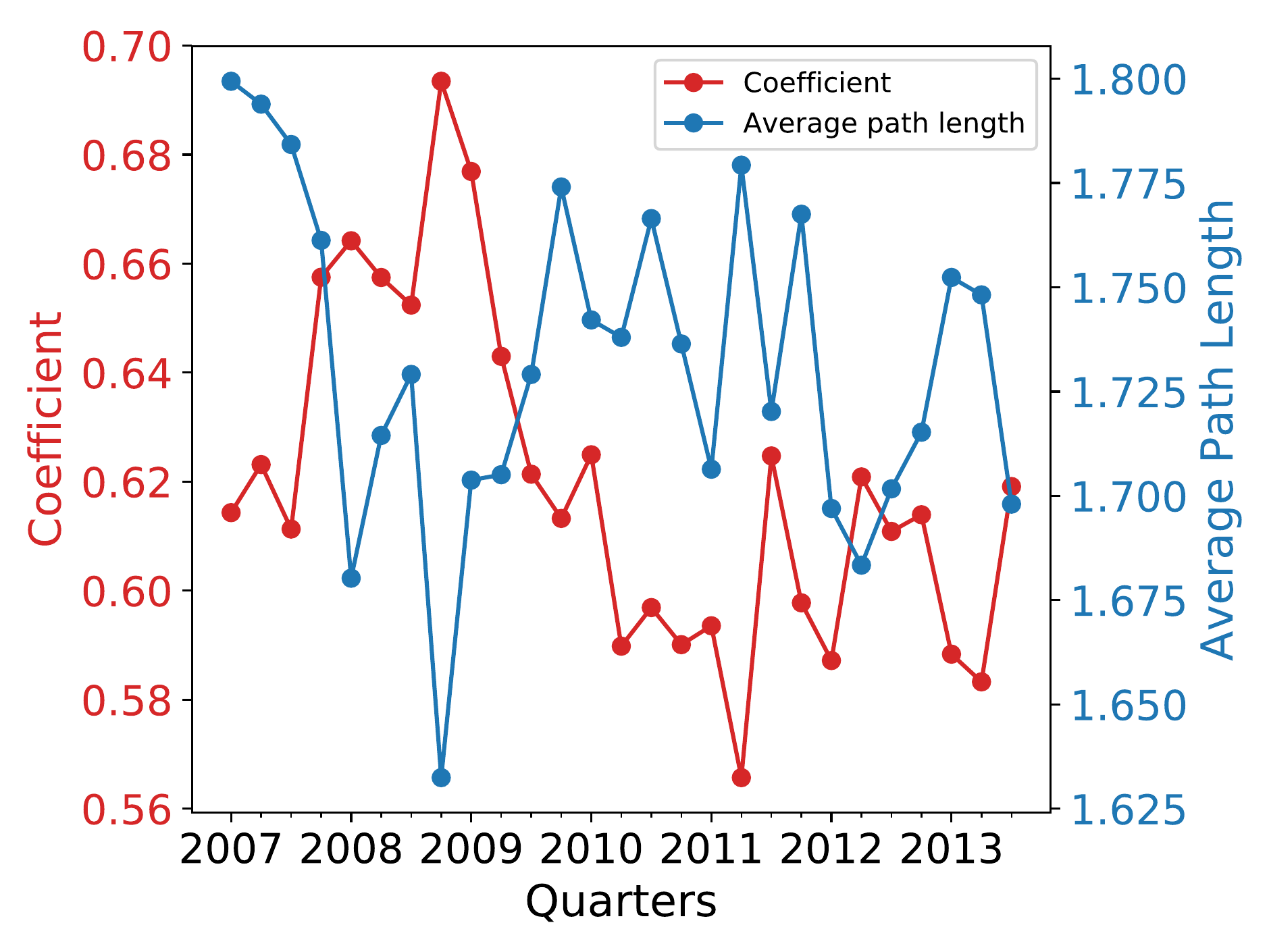}}
  \subfloat[]{\includegraphics[width =0.3\textwidth]{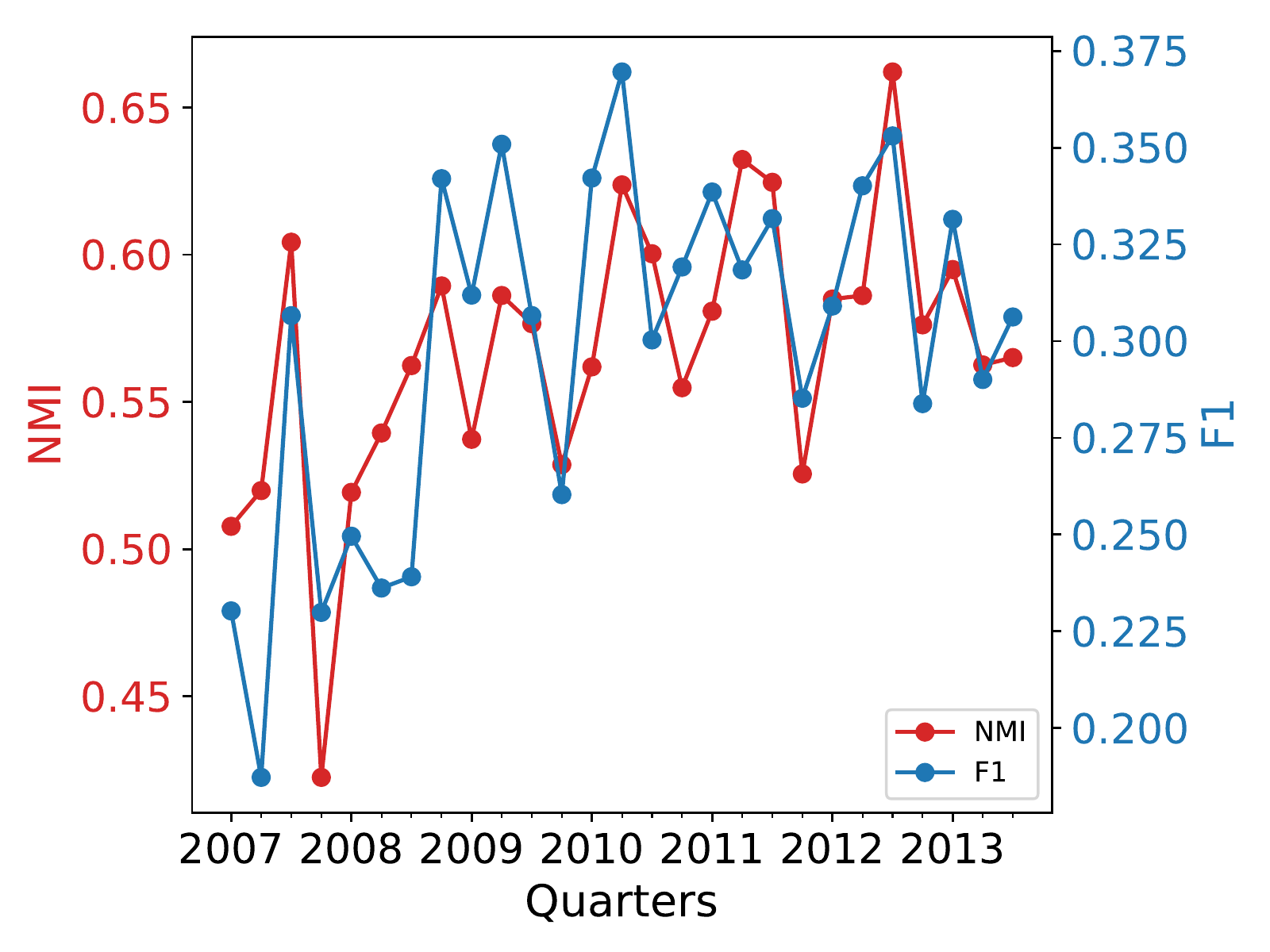}}

  \caption{Evolution of news co-occurrence network features and dynamic comparison with ground-truth sectors. (a) Average degree; (b) Clustering coefficient and average path length; (c) Auto-detected groups vs. ground-truth sectors.}
  \label{fig:quarter_network}
\end{figure}

\clearpage
\floatsetup[figure]{subcapbesideposition=top}
\begin{figure}[t]
   \subfloat[]{\includegraphics[width =0.20\textwidth]{newfigures/group_px/GM_pos.png}}
   \subfloat[]{\includegraphics[width =0.20\textwidth]{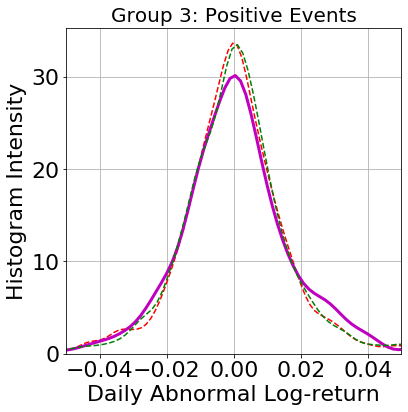}}
   \subfloat[]{\includegraphics[width =0.20\textwidth]{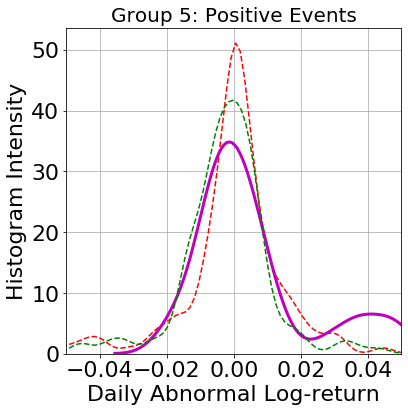}}
   \subfloat[]{\includegraphics[width =0.20\textwidth]{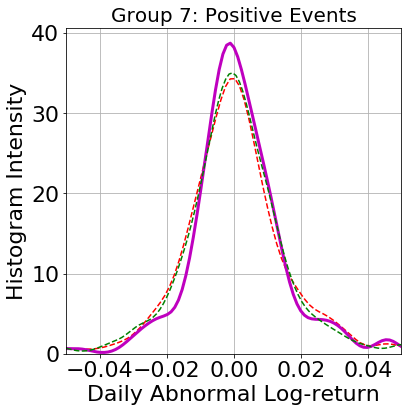}}
   
      \includegraphics[width = 0.4\textwidth]{newfigures/legend_group.pdf}

  \caption{Probability density function of \textbf{AR} of companies in several groups, computed over all group-aggregated \textbf{positive} sentiment events over the time period: (a) Group 2, (b) Group 3, (c) Group 5, (d) Group 7.}

\label{fig:pxgrouppositive}
\end{figure}

\clearpage
\floatsetup[figure]{subcapbesideposition=top}
\begin{figure}[t]
   \subfloat[]{\includegraphics[width =0.20\textwidth]{newfigures/group_px/GM_neg.png}}
   \subfloat[]{\includegraphics[width =0.20\textwidth]{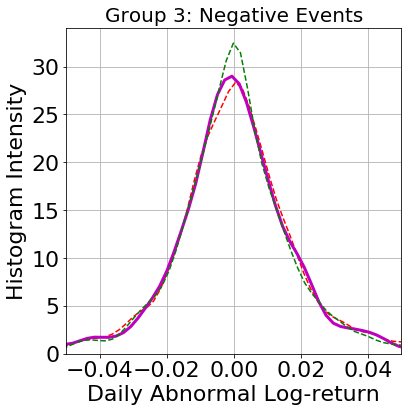}}
   \subfloat[]{\includegraphics[width =0.20\textwidth]{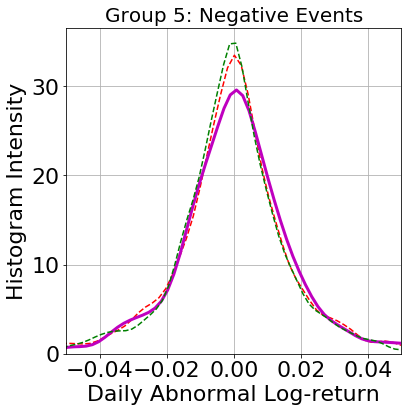}}
   \subfloat[]{\includegraphics[width =0.20\textwidth]{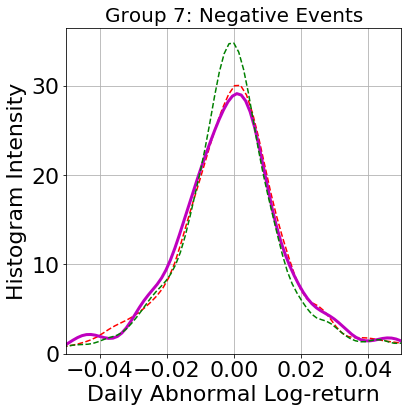}}
  
    \includegraphics[width = 0.4\textwidth]{newfigures/legend_group.pdf}
  \caption{Probability density function of \textbf{AR} of companies in several groups, computed over all group-aggregated \textbf{negative} sentiment events over the time period: (a) Group 2, (b) Group 3, (c) Group 5, (d) Group 7. }

\label{fig:pxgroupnegative}
\end{figure}

\clearpage
\floatsetup[figure]{subcapbesideposition=top}
\begin{figure}[t]
   \subfloat[]{\includegraphics[width =0.20\textwidth]{newfigures/group_vol/GM_pos.png}}
   \subfloat[]{\includegraphics[width =0.20\textwidth]{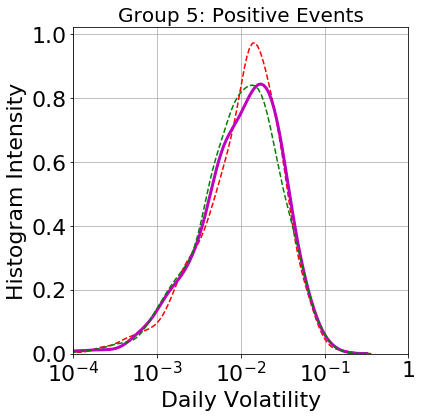}}
   \subfloat[]{\includegraphics[width =0.20\textwidth]{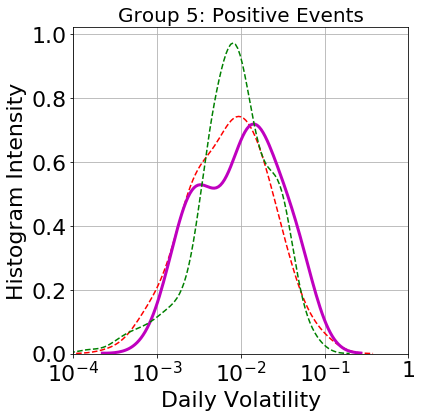}}
   \subfloat[]{\includegraphics[width =0.20\textwidth]{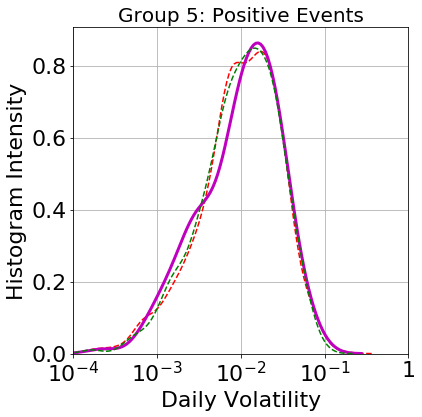}}
  
    \includegraphics[width = 0.4\textwidth]{newfigures/legend_group.pdf}
  \caption{Probability density function of \textbf{Daily realised volatility} of companies in several groups, computed over all group-aggregated \textbf{positive} sentiment events over the time period: (a) Group 2, (b) Group 3, (c) Group 5, (d) Group 7.}

\label{fig:volgrouppositive}
\end{figure}

\clearpage
\floatsetup[figure]{subcapbesideposition=top}
\begin{figure}[t]
   \subfloat[]{\includegraphics[width =0.20\textwidth]{newfigures/group_vol/GM_neg.png}}
   \subfloat[]{\includegraphics[width =0.20\textwidth]{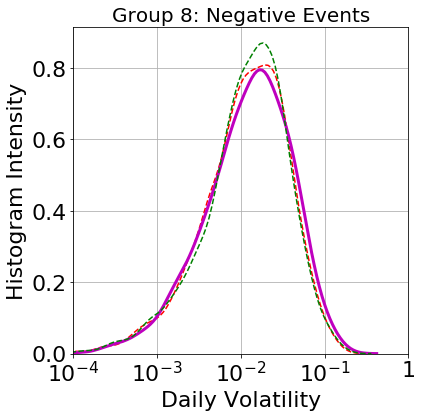}}
   \subfloat[]{\includegraphics[width =0.20\textwidth]{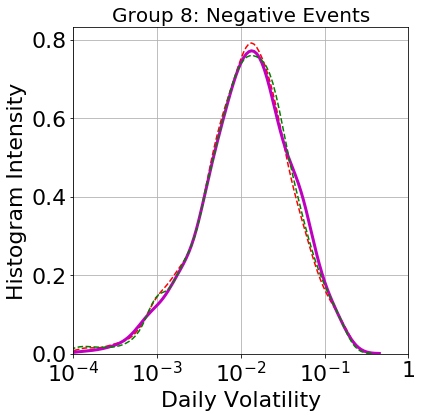}}
   \subfloat[]{\includegraphics[width =0.20\textwidth]{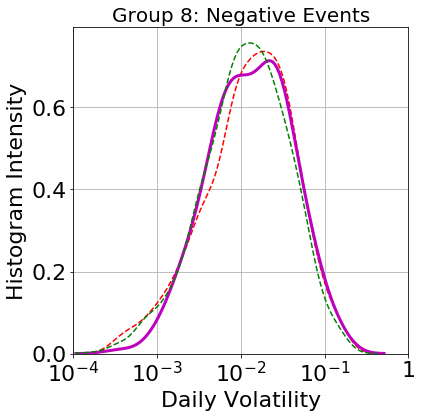}}
  
    \includegraphics[width = 0.4\textwidth]{newfigures/legend_group.pdf}
  \caption{Probability density function of \textbf{Daily realised volatility} of companies in several groups, computed over all group-aggregated \textbf{negative} sentiment events over the time period: (a) Group 2, (b) Group 3, (c) Group 5, (d) Group 7.}

\label{fig:volgroupnegative}
\end{figure}

\end{document}